\documentclass[twocolumn,showpacs,showkeys,preprintnumbers]{revtex4}
\usepackage{graphicx}
\usepackage{subfigure}
\usepackage{lscape}
\usepackage{dcolumn}
\usepackage{bm}
\usepackage{color}

\usepackage{amsmath} 
\usepackage{amssymb} 
\usepackage{multirow}
\usepackage{epstopdf}

\begin{document}
\title{The shape of the electron and muon lateral distribution functions of extensive air showers}
\author{A Basak}
\email{ab.astrophysics@rediffmail.com}
\author{R K Dey}
\email{rkdey2007phy@rediffmail.com}
\affiliation{Department of Physics, University of North Bengal, Siliguri, WB 734 013 India}

\begin{abstract}
The lateral density data obtained for different secondaries of an extensive air shower (EAS) from an array of detectors are usually described by some suitable lateral density functions (LDFs). Analyzing non-vertical simulated EASs generated with the CORSIKA code, it is found that the lateral and polar density distributions of electrons and muons are asymmetric in the ground plane. It means that typical expressions for symmetric lateral density functions (SLDFs) (\emph{e.g.} the Nishimura-Kamata-Greisen function) are inadequate to reconstruct the lateral and polar dependencies of such asymmetric electron or muon densities accurately.  In order to provide a more consistent LDF for non-vertical shower reconstruction in the ground plane, the paper considers the issue of the modification of the SLDF analytically. The asymmetry arising from additional attenuation and correction of the positional coordinates (radial and polar) of cascade particles causes a gap length between the center of concentric equidensity ellipses and the EAS core. A toy function is introduced as a basic LDF to describe the asymmetric lateral and polar density distributions of electrons or muons of EASs, thereby predicting the gap length parameter.  Consequently, the desired LDF describing the asymmetric density distributions of electrons and muons of EASs has emerged. We compare results from detailed simulations with the predictions of the analytical parametrization. The LDF derived in this work is found to be well-suited to reconstruct EASs in the ground plane directly.   

\end{abstract}

\maketitle

\section{Introduction}	
Secondary cosmic ray (CR) particles of an extensive air shower (EAS) advance towards the ground as a thin disk through the atmosphere from the direction of their parent primary CR particle at the speed of light. After the first interaction point, somewhere below the top of the atmosphere, the disk begins to form, continues to grow, and then starts attenuating after the depth of shower maximum. The transverse and longitudinal momenta imparted on the shower particles emerging from their parent particles via the hadronic interactions would cause the lateral and longitudinal spreads for these particles in an EAS [1].

In various EAS data analyses, a shower is approximated to a cylinder over a finite height before arriving at the ground. In this cylinder model, all the particles in the EAS are assumed to advance parallel with the EAS core, and equidensity contours of particle densities are also considered as circles in the shower plane. For a vertical shower, the equidensity contours of the particle densities are generally treated as circles in the observation plane on the ground, as the shower plane of a vertical shower coincides with the observation plane. A symmetric lateral density function (SLDF) such as the well-known Nishimura-Kamata-Greisen (NKG) type LDF can convincingly describe the lateral density distributions (LDD) of shower particles of vertical showers [2-3].

For non-vertical showers with increasing zenith angle, the equidensity contour of the particle density configuration in the observation plane on the ground changes more and more from a circular to an elliptical shape, chiefly due to the so-called geometric effect [4-5]. At the same distance $r_{g}$ from the EAS core in the observation plane on ground, the effect will enhance particle densities for the late and advance regions of a shower corresponding to polar angles, $\beta_{g}=0^{o}$ and $\beta_{g}=180^{o}$ compare to the mutual/intersecting regions for $\beta_{g}=90^{o}$ and $\beta_{g}=270^{o}$. The polar angle is taken in an anti-clockwise sense to the positive x-axis in the observation plane, $g$-plane, as well as in the shower plane, $s$-plane (see Fig. 1c). Geometric effect unveils one more feature (known as the $\cos{\Theta}$ effect), where the particle density at the radial distance $r_g$ in the observation plane coincides to the particle density at the radial distance $r_{s}=r_{g}\cos{\Theta}$ for $\beta_{s}=0^{o}$ and $\beta_{s}=180^{o}$ in the shower plane. However, these two densities match each other for the same $r=r_{g}=r_{s}$ in both the planes corresponding to $\beta=90^{o}$ and $\beta=270^{o}$.

Entangled with the geometric effect, the varying atmospheric column densities traversed by the EAS particles at different locations of the EAS will further enhance the polar asymmetry in the LDD data. This feature of an EAS is categorized as an attenuation effect on the asymmetry of particle densities [4-5]. This additional effect causes a reasonable density asymmetry between the late region ($\beta_{g}=0^{o}$) and the early region ($\beta_{g}=180^{o}$) of an inclined EAS. Particles reaching well advanced at the observation plane suffer less atmospheric attenuation than particles arriving late as they have traveled longer paths. The EAS particles striking all the other points in the observation plane except those populated along the boundary across $\beta_{g}: 90^{o} - 270^{o}$, separating the late and the early regions would give rise to an asymmetry in particle density due to the effect. 

The influence of Earth's magnetic field on the LDD of charged particles would lead to another asymmetry effect. In a continuously developing cascade, the charged electromagnetic (EM) component, i.e. electrons ($e$) (henceforth $e = e^{+}+e^{-}$), possess shorter radiation lengths participating in dominant EM processes in the atmosphere. They experience intense radiative losses via bremsstrahlung processes,  moving arbitrarily relative to the geomagnetic field. All these strongly interacting processes of electrons allow them to form a wider lateral spread but restrain them from reasonable geomagnetic influences. Hence, their LDDs give rise to minimal asymmetries caused by the geomagnetic field. Therefore, the asymmetry in the LDDs of electrons in the observation plane on the ground mainly arises from a combination of geometric and attenuation effects. In contrast to the behaviour of electrons in the geomagnetic field, muons travel longer distances in the atmosphere with less probable EM and very negligible weak interaction processes. Thus, muons experience the influence of the geomagnetic field for a longer duration, thereby contributing asymmetries to their LDDs, even for nearly vertical showers [4-5]. 

Several studies have been carried out on modeling the geometric and attenuation effects for non-vertical showers in the form of giving an accurate/refined LDF by adopting the cylinder shower model in recent papers (see, for instance, [4-6] and references therein). In practice, however, the shower periphery varies with the advancement of an EAS. First, it rises and later starts shrinking after the shower maximum. Thus, the distances of equidensity contours from the EAS axis start contracting (density contours are still assumed as shrinking circles) after the shower maximum in the shower plane. This situation is analogous to an inclined, inverted truncated cone. The landing base of the cone on the observation plane is composed of equidensity contours having an elliptical shape. The main focus of the work is to construct a more accurate LDF for describing the density of the particles in the observation plane from an assumed and simplified polar symmetric density in the shower plane (i.e. from an NKG-type SLDF) based on the \emph{cone model}. 

We have understood that the plane of the shower front hitting the observation level is a collection of equidensity circles for vertical showers, and the centre of these concentric circles coincides with the EAS core. In the case of non-vertical showers, however, the projection of the shower plane, hitting the observation plane on the ground, instead consists of equidensity ellipses, for which their centers do not meet with the core of the shower. A linear distance results between the centre of an arbitrary elliptic density contour and the EAS core, and is named the gap length, $x_C$ (see Fig. 1a). The varying attenuation that the EM and muonic components encounter in the late and early regions of the shower front to the ground plane forms $x_C$ and thereby contributes to reasonable polar density variation. The inaccurate SLDF for describing the polar- and lateral-dependent densities of shower $e$/muons (henceforth $\mu=\mu^{+}+\mu^{-}$) for non-vertical showers needs to be corrected by incorporating $x_C$ parameter into it, and is named as the elliptic-LDF (ELDF). For non-vertical showers with zenith angle ($\Theta \ge 40^o$), $x_C$ has emerged as an essential parameter sensitive to the nature of the shower initiating CR particle and an anchor for the proposed ELDF analytically. The ELDF facilitates a more accurate shower data analysis for reconstructing lateral/polar profiles of EASs on the ground/detector plane and, therefore, of the universal primary CR (PCR) mass-sensitive EAS observables, such as the shower size ($N_e$), muon size ($N_{\mu}$), lateral shower age ($s_{\perp}$) [7], local shower age ($s_{\emph{local}}$) [8], and others.

The earlier efforts used a cylinder model for the evolution of the EAS in the atmosphere to shape such an ELDF [6]. The feature of shrinking equidensity contours in the cascade development has been ignored in the parametrization of the ELDF based on the cylinder shower model. Consequently, the additional atmospheric depth encountered by EAS particles between the shower and ground planes has been calculated from apex $P$ for the cone model (see Fig. 1c). However, in the cylinder model, such an additional atmospheric depth parallel to the shower axis was accounted for just by taking a difference between a point in the shower plane and its corresponding point projected on the observation plane in the ground. The revised path length obtained in the cone model will give rise to a different measure of the attenuation of the EAS particles between the planes. Hence, we shape the ELDF here by including $x_C$ into the SLDF based on the \emph{cone model} for the EAS, which can be applied directly to the LDD data obtained from the detector plane. To that end, the modeling of the atmospheric attenuation that mainly causes $x_C$ will be described. In earlier works, the refined LDD of muons of highly inclined showers was undertaken in EAS-data analysis focusing on some particular aspects. Hence, the corresponding LDFs of the LDD of muons were advocated [5,9-10]. These papers mainly dealt with the effect of the scheme of transforming the LDD of muons from the detector plane to the shower plane with a focus on analyzing various EAS observables, thereby exploring the nature of the EAS-initiating CR particle.

The rest of the paper is structured as follows. In section 2, we will analytically present the main features of the \emph{cone model} of shower development. In the same section, the parametrization for the shift of the EAS core and the ELDF of the EAS particles will be discussed. Section 3 discusses the attributes of the Monte Carlo (MC) simulations. The method for analyzing MC data to estimate the $x_C$ parameter is described in Section 4. We then present our results and pertinent discussions in sections 5 and 6. Finally, section 7 provides a summary and conclusion.

\section{Scheme of the EAS geometry}
\subsection{Cone model}
We have realized that the equidensity contours of EAS $e/\mu$-s in the shower plane are circular. The effective area of the observation plane on the ground depends upon the energy ($E$) and zenith angle ($\Theta$) of the CR particle [11]. The density and timing data of $e/\mu$-s obtained from particle detectors positioned in the observation plane of an EAS array have been used to estimate the CR energy and zenith angle. 

The sketch in the Fig. 1c of a shower shows different paths from the apex $P(x_p,y_p,z_p)$ across the layers of the atmosphere and extending up to the observation plane on the ground. The x-y plane of the coordinate system describes the observation plane, and the z-axis manages the vertically upward direction. The point of intersection between the shower axis and the observation plane is treated as the origin of the coordinate system. The shower azimuth angle $\Phi$ is taken between the positive x-axis and the horizontal component of the momentum vector of the shower-initiating particle following a counter-clockwise sense. $\Theta$ is measured between the shower axis and the negative z-axis. A schematic view of a shower evolution in the atmosphere is shown in Fig. 1b.

\begin{figure}[htbp]
 \centering
  \includegraphics[trim=1.2cm 0.9cm 0.9cm 0.9, clip=true, totalheight=0.26\textheight,angle=0]{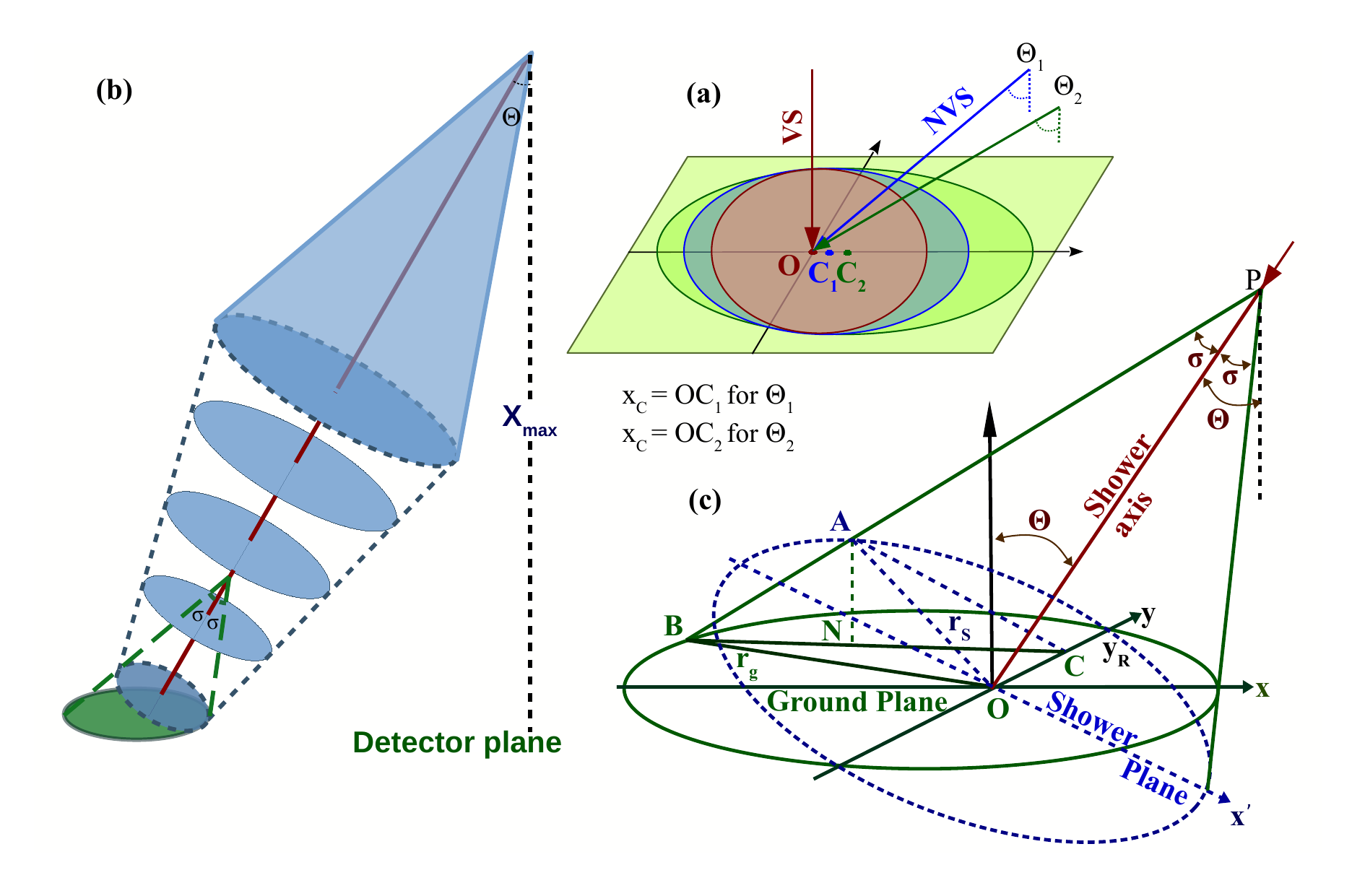}
   \caption{The basic sketch of a shower falling on an observation plane in the ground surface. Shower plane: dotted circle; Observation plane: solid ellipse in Fig. 1b and c. Fig. 1a displays equidensity contours on the observation plane for one vertical shower (VS) and two non-vertical showers (NVS). The EAS core and centre of the equidensity contour (circle) coincide for VS. For NVS, the EAS core and the centre of the equidensity contour (ellipse) do not overlap. Here, $x_{c1}$ and $x_{c2}$ deal with the gap lengths for two different showers with zenith angle, $\Theta_{1}$ and $\Theta_{2}$.}
\end{figure}

We incorporate non-vertical showers approaching from the North in the current modeling of the geometry and attenuation effects for showers (i.e. $\Phi = 0 ^o$). The tilted circle with a dotted boundary represents the shower plane perpendicular to the shower axis. We refer to a point $A$ with coordinates $(x_s,y_s,z_s)$ on the tilted circle as shown in Fig. 1c. Distance of $A$ from the origin $O$ is taken as $r_s$, whereas $r_g$ accounts for the distance of the point $B$ from $O$. Here, point $B(x_g,y_g,z_g)$ is the projection of $A$, along the slant height of the cone, onto the observation plane. Distances of these points $A$ and $B$ from $O$ along the y-direction are equal and given by the element OC $(OC=y_s=y_g$; $C$ lies along the y-axis where the shower plane and the ground plane intersect each other). Line elements AC and BC make an angle equal to $\Theta$ of the EAS. Now, we can relate all these line elements from the geometry to obtain the following connections:
\begin{equation}
  \begin{aligned}
    & AO^2 = OC^2 + AC^2 \implies 
    r_s^2 = y_s^2 + x_s^2 
    \\ & AC^2=r_s^2 - y_g^2
  \end{aligned}
\end{equation}
Let's look at the enlarged view of Fig. 1c via Fig. 2 to understand all features of the geometric effect clearly.
\begin{figure}[htbp]
 \centering
  \includegraphics[trim=0.0cm 1.0cm 0.0cm 0.0cm, scale=0.4]{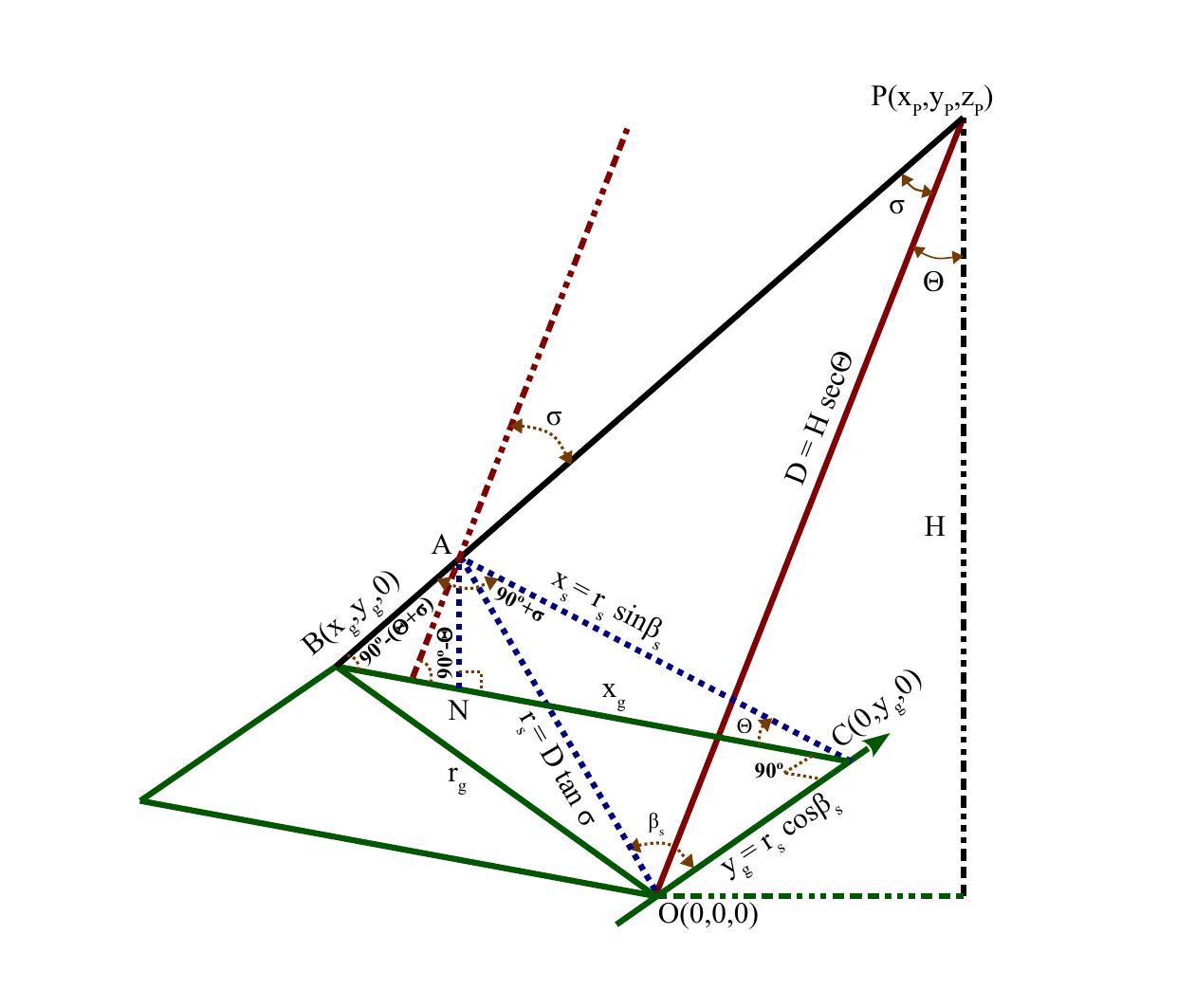}
   \caption{Enlarged view of Fig. 1c in the EAS intrinsic coordinate system.}
\end{figure}

Using the law of sines in the $\bigtriangleup ABC$, we obtain,
\begin{equation*}
  \frac{AB}{\sin\Theta}=\frac{BC}{\sin(90+\sigma)}=
    \frac{AC}{\sin(90-(\Theta+\sigma))},
\end{equation*}
Now, 
 $BC = x_g$,
 $AC = x_s $ from Fig. 2, then
\begin{equation*}
 \frac{AB}{\sin\Theta}=\frac{x_g}{\cos\sigma}=\frac{x_s}{\cos(\Theta+\sigma)}  \implies AB = \frac{x_g\sin\Theta}{\cos\sigma},
\end{equation*}
\begin{equation}
  x_s = x_g \frac{\cos(\Theta + \sigma)}{\cos \sigma}
\end{equation}
Putting Eq. (2) and also $y_s=y_g$ in Eq. (1), we get,
\begin{equation}
  r_s^2 = y_g^2 +  x_g^2\frac{\cos^2(\Theta + \sigma)}{\cos^2\sigma}
\end{equation}

It can be seen from Fig. 1c that if the attenuation process in the region between the observation and shower planes is ignored, then these two planes intersect each other along the positive y-axis at $y_R$, arising out of the geometric effect only. With the distance $y_R=r_s$ in Eq.~(3), the corresponding equation works in the observation plane on the ground is,
\begin{equation}
  x_g^2\frac{\cos^2(\Theta + \sigma)}{\cos^2\sigma} + y_g^2 = y_R^2
\end{equation}  
Eq.~(4) represents an ellipse where the EAS core is located at its centre.  
It was already stated in Sec. I that the atmospheric attenuation of EAS particles causes a linear shift of the centre of the ellipse from the EAS core. For $\Phi = 0^o$, the shift will occur only along the x-direction, and we define the x-coordinate of the shifted centre as $x_C$ in the observation plane. With the shifted centre, the equation of the ellipse becomes
\begin{equation}
  (x_g-x_C)^2\frac{\cos^2(\Theta + \sigma)}{\cos^2\sigma} + y_g^2 = b^2,
\end{equation}	
, where \textit{b} is the semi-minor axis length for the shifted ellipse. Here, if we set $x_g=0$, then distances $y_g$ and $y_R$ are equal, we obtain 
\begin{equation}
  b^2 = x_C^2\frac{\cos^2(\Theta + \sigma)}{\cos^2\sigma} + y_R^2 
\end{equation}
Finally, inserting Eq.~(6) into Eq.~(5), we will find the governing equation for the modified ellipse as 
\begin{equation}
  (x_g^2-2 x_g x_C)\cdot \frac{\cos^2(\Theta + \sigma)}{\cos^2\sigma} + y_g^2 = y_R^2
\end{equation}
To have an ELDF for an EAS that experiences both geometric and attenuation effects, the solution to Eq.~(7) is crucial.

\subsection{Method for modeling the attenuation of electrons/muons in an EAS}
When attenuation is absent, the lateral density of EAS particles ($\rho_{g;e/\mu}$) in the observation plane on the ground coincides with the same in the shower plane ($\rho_{s;e/\mu}$) for vertical showers but differs for inclined showers, and are connected via a simple geometric transformation,
\begin{equation}
  \rho_{s;e/\mu}(r_s)  = \frac{\rho_g(r_{g;e/\mu})}{ \cos\Theta}.
\end{equation}
The magnitude of the ${N_{e}}~{\text{or}}~{N_{\mu}}$ size first increases as an EAS travels through the atmosphere because shower secondaries are produced faster than they are attenuated. However, as it approaches the depth of the shower maximum ($X_{max}$), it gradually declines because of the counter effect. Fig. 1 indicates that when the depth of the shower surpasses $X_{max}$, an EAS undergoes its attenuation phase. Due to attenuation occurring while moving from point A to point B, the density of shower particles decreases exponentially by a factor of $e^{-\Delta X /\Lambda}$ [1-2]. Here, $\Delta X =(X-X_g)$ ~g~cm$^{-2}$ measures the extra path traversed by EAS particles from A to B, and $\Lambda$ is the attenuation length in g~cm$^{-2}$. The value of $\Lambda$ depends on the kind of EAS secondary particles that are attenuated in the atmosphere. In the observation plane on the ground, the attenuated density $\rho_{g;e/\mu}$ of a specific type of EAS particle is given by 

\begin{equation}
  \rho_{g;e/\mu}(r_g)= \cos\Theta \cdot \rho_{s;e/\mu}(r_s) \cdot e^{-\frac{\Delta X}{\Lambda_{e/\mu}}}
\end{equation}
Now, the coordinate of the apex P, according to Fig. 2, is		
\begin{equation*}
\begin{aligned}
       (x_P,y_P,z_P)\equiv (D\sin\Theta,0,D\cos\Theta)
\end{aligned}
\end{equation*}
From Eq.~(2), we obtain
\begin{equation*}
\begin{aligned}
	x_g &= x_s\frac{\cos\sigma}{\cos(\Theta+\sigma)} 
	= \frac{D\tan\sigma \sin\beta_s}{\cos\Theta (1-\tan\Theta \tan\sigma)}
\end{aligned}
\end{equation*}	
where, the Fig. 2 provides~
	$x_s = r_s \sin \beta_s$ and $ r_s =  D ~\tan \sigma$.\\\\
From the same figure, we have 
\begin{equation*}
	y_g =  r_s \cos\beta_s = D \tan\sigma \cos\beta_s
\end{equation*}	
As a result, the coordinates of point B are, 		
\begin{equation*}
	(x_g,y_g,z_g)\equiv \Big(\frac{-~D\tan\sigma\sin\beta_s}{\cos\Theta(1-\tan\Theta\tan\sigma)}, D\tan\sigma\cos\beta_s,0 \Big)
\end{equation*}
By only the coordinates as mentioned above and taking only the first order terms in $\tan\sigma$, we can calculate the atmospheric depth along the path length PB. 

\begin{equation}
\begin{aligned}
	\overline{PB} &= D \Big[ \Big(\sin\Theta+\frac{\tan\sigma\sin\beta_s}{\cos\Theta(1-\tan\Theta\tan\sigma)}\Big)^2 
	\\ &+\tan^2\sigma\cos^2\beta_s+\cos^2\Theta \Big]^{1/2}
	\\ &\approx D\Big(1+\frac{\tan\sigma\sin\beta_s\tan\Theta}{1-\tan\Theta\tan\sigma}\Big)	
\end{aligned}
\end{equation} 
On the other hand, the slant depth along the path PO is 
\begin{equation}
	\overline{PO}=D=H\sec\Theta
\end{equation}
 where $H$ accounts for the height of the apex of the cone. Based on the Fig. 2, we can derive the following from geometry:
\begin{equation}
	\tan\sigma = \frac{r_{s}}{D}=\frac{r_{s}}{H\sec\Theta}
	\implies
	\tan\Theta \tan\sigma =\frac{r_{s} \sin \Theta}{H} 
\end{equation}
For example, the atmospheric depth is 1022 g~cm$^{-2}$, and the slant depth is $1022 \cdot \sec \Theta$ ~g~cm$^{-2}$ at the KASKADE site [12].   

In general, the difference $\Delta X$ in atmospheric depth between the paths $\overline{PB}$ and $\overline{PO}$ is 
\begin{equation}
	\Delta X=\frac{D\sin\beta_{s} \tan\Theta \tan\sigma}{1-\tan\Theta \tan\sigma}
\end{equation}	
One meter traversal by EAS particles at the KASCADE experiment site corresponds to $c_{f}\approx 0.15$~ g~cm$^{-2}$ atmospheric depth based on the atmospheric composition there [12-14]. As a result, the extra path travelled by the EAS particles in linear and density scales  can be equated as follows,
\begin{equation*}
	\eta \cdot \overline{AB} = \Delta X /\Lambda \implies \eta = (\frac{\Delta X}{AB}) \cdot \frac{1}{\Lambda} = \frac{c_f}{ \Lambda },
\end{equation*}
where $\eta$ is the attenuation length in units of the reciprocal of linear distance. 
In the above expression, we now insert the formula of $\Delta X$ from Eq.~(13) and also the final relationship in Eq.~(12); we then get the following,
\begin{equation*}
\begin{aligned}
	\eta \cdot AB &=\frac{\eta{H}}{c_f} \cdot \frac{x_g \tan \Theta\cos(\Theta + \sigma)}{\cos\sigma{(H-r_s \sin \Theta)}}
\end{aligned}
\end{equation*}	
Eq.~(9) is reduced to the following form by converting units to a linear scale: 
\begin{equation}
	\rho_g(r_g)= \cos\Theta \cdot \rho_s(r_s) \cdot e^{-\eta \cdot AB} 
\end{equation} 
Based on the present modeling for the shower evolution geometry, we have to substitute $[\pm \frac{H{x_g}  \tan \Theta}{c_f} \cdot \frac{ \cos(\Theta + \sigma)}{\cos\sigma\cdot (H-r_s \sin \Theta)}]$ for $AB$. In Eq.~(14), the negative sign for the additional path $AB$ corresponds to the attenuation of the delayed part of the EAS. We subsequently derive,
\begin{equation}
	\rho_g(r_g) = \cos\Theta \cdot \rho_s(r_s) \cdot e^{\frac{\eta{H}x_g \tan \Theta}{c_f} \cdot \frac{\cos(\Theta + \sigma)}{\cos\sigma\cdot (H-r_s \sin \Theta)}}
\end{equation}
Eq.~(4) yields $y_g = y_R$ while $x_g$ takes $0$ and the following is what Eq.~(15) simplifies to 
\begin{equation}
	\rho_g(r_g)_{x_g=0,y_g=y_R} = \cos\Theta \cdot \rho_s(y_R) 
\end{equation}
For the specific case, $y=y_R$, the density of particles in an elliptic contour in the observation plane on the ground will be free from $x_g$. It then offers the equality as $\rho_g(x_g,y_g) = \rho_g(0,y_R)$ and hence Eq.~(16) can be rewritten as
\begin{equation}
	\rho_g(r_g / (x_g,y_g)) = \cos\Theta \cdot \rho_s(y_R). 
\end{equation}
We derive one of the governing equations dealing with attenuation outcomes on the density of EAS particles by substituting Eq.~(17) into Eq.~(15) as follows, 
\begin{equation}
	\rho_s(y_R) =  \rho_s(r_s) \cdot e^{\frac{\eta{H} x_g  \tan \Theta}{c_f} \cdot \frac{\cos(\Theta + \sigma)}{\cos\sigma\cdot (H-r_s \sin \Theta)}}
\end{equation}	

\begin{figure}[!h]
	\centering
	\includegraphics[trim=0.0cm 0.0cm 0.0cm 0.0cm, scale=0.30]{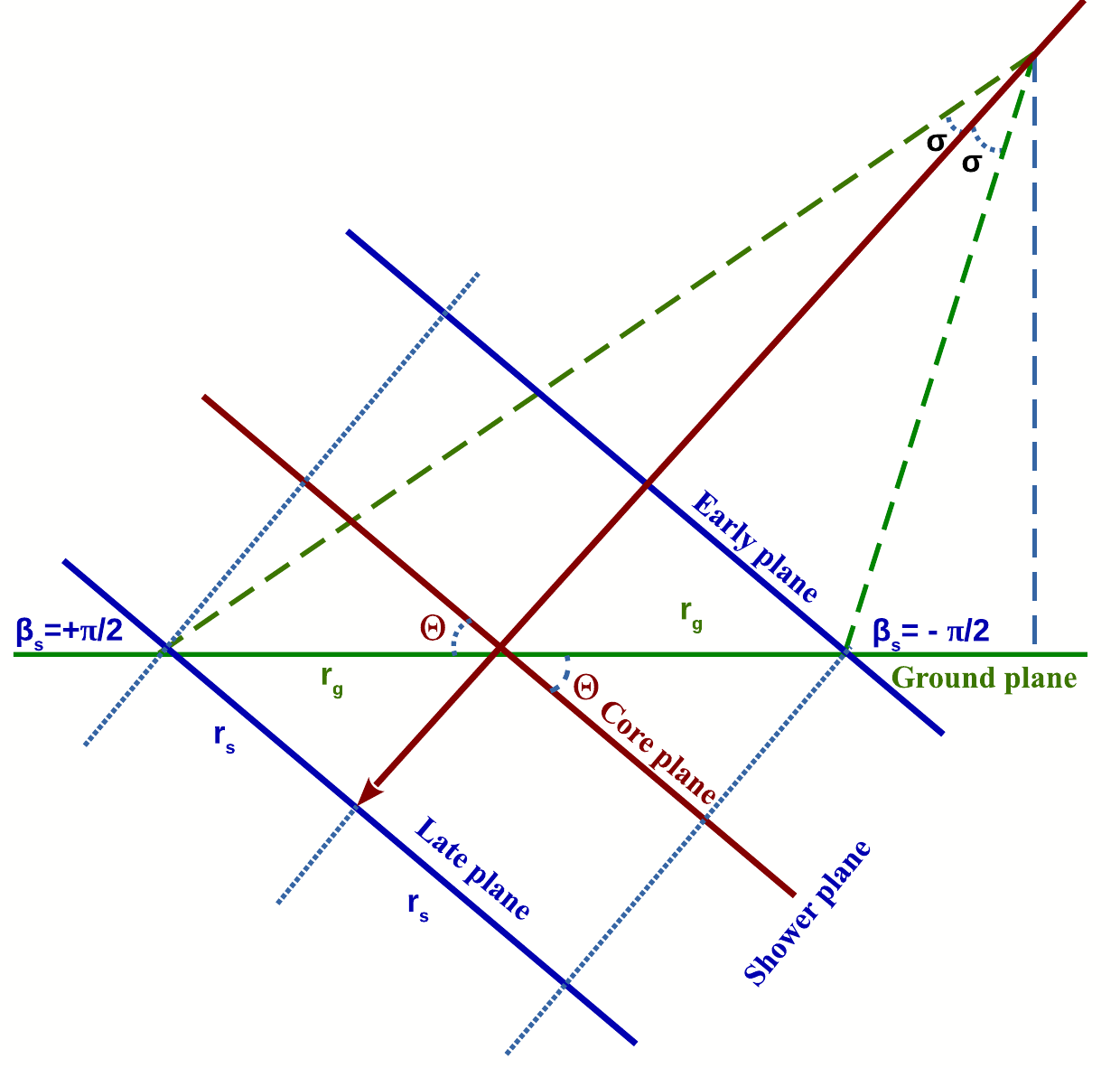}
	\caption{The shower plane intersects the observation plane on the ground at three different depths of the EAS development from the apex.}
\end{figure}
Fig. 3 indicates that the EAS particles arriving at the ground in the early region of the shower at $\beta_s = -\pi/2 $ have less inclination than those in the late region at $\beta_s = \pi/2 $. The overall extra path traversed by the particles in the late region compared to the early region of an EAS can be obtained by using Eq.~ (13) as follows, 
\begin{equation*}
	2 \times  \Delta X \approx  H \frac{ 2 \times \tan\Theta \tan\sigma }{\cos \Theta(1-\tan\Theta \tan\sigma)}.	
\end{equation*}
Finally, the height of the apex $H$ of the cone can be given as below from the above expression after replacing the term $\tan\Theta \tan\sigma$ by $r_s\sin\Theta/H$,
\begin{equation}
	H = \frac{r_s \sin\Theta}{(1+\frac{r_s~\tan\Theta}{\Delta{X}}).}  
\end{equation}
 
\subsection{Estimating the gap length parameter using a toy function for LDD of EAS particles}
There were a wide variety of LDFs, namely the NKG [3], Uchaikin  [15], Linsley [16], Hillas [17], and some modified NKG forms [7, 18-21] were exploited by different EAS experiments to fit the simulated/observed LDDs of $e/\mu$-s. In modeling the ELDF, we must move forward from Eq.~(18) to obtain first an expression for the gap length parameter, $x_C$ and finally, the required ELDF in a simplified and more appropriate manner. Moreover, we knew that some earlier EAS experiments, such as Haverah Park [17], MSU [21] and Tien-Shan [22], applied an exponential LDF to the LDDs of electrons/muons or occasionally for hadrons. Keeping all these in mind, and after testifying the applicability of the exponential LDF to our simulated LDDs of $e/\mu$-s, we considered the LDF as a \emph{toy function} (TF) to work out the method. The intended TF is listed below:

\begin{equation}
	\rho_{s}(r_s) \simeq c \cdot e^{-\alpha (\frac{r_s}{r_0})^\kappa},
\end{equation}

where $c$ is a constant expressed in m$^{-2}$, and $\alpha$ and $\kappa$ are dimensionless. The trio $c$, $\alpha$ and $\kappa$ depend upon the nature of the primary and secondary CR particles and the high-energy hadronic interaction models. Through Table III, we have shown these dependencies. The following can be written once we include the TF in Eq.~(18).  
\begin{equation}
	e^{-\alpha (\frac{y_R}{r_0})^\kappa} =e^{\frac{\eta{H} x_g  \tan \Theta}{c_f} \cdot \frac{\cos(\Theta + \sigma)}{\cos\sigma\cdot (H-r_s \sin \Theta)}} \cdot  e^{-\alpha (\frac{r_s}{r_0})^\kappa}
\end{equation}
Then, we get the following
\begin{equation}
	r_s = y_R \cdot \Bigg[1+{  \frac{\eta{H}x_g  \tan \Theta ~\cdot~\cos(\Theta + \sigma)}{\alpha(\frac{y_R}{r0})^\kappa\cos\sigma\cdot (H-r_s \sin \Theta)}}\Bigg]^{1/\kappa}
\end{equation}
After squaring and rearranging Eq.~(22), we may write, 
\begin{equation}
	r_s^2 = y_R^2 \cdot \Bigg[1+{ \frac{  x_g\kappa{H} \cos(\Theta + \sigma)}{{c_f}\cos\sigma\cdot (H-r_s \sin \Theta)} \cdot \frac{\eta \tan \Theta}{\alpha\kappa(\frac{y_R}{r0})^\kappa}}\Bigg]^{2/\kappa}
\end{equation}
Using $\delta$ in place of $\frac{\eta \tan \Theta}{\alpha\kappa(\frac{y_R}{r0})^\kappa}$, Eq.~(23) can be rewritten as,
\begin{equation}
	r_s^2 = y_R^2 \cdot \Bigg[1+{ \frac{{H}x_g\kappa\delta  \cos(\Theta + \sigma)}{{c_f}\cos\sigma(H-r_s \sin \Theta)}}\Bigg]^{2/\kappa}
\end{equation}
The second term on the right-hand side of the above equation is substantially smaller than 1 for square-shaped EAS arrays with $x_g \leq 10^3$~m, and we will implement a first-order approximation in its expansion. Hence,
\begin{equation}
	r_s^2 \approx y_R^2 \cdot \Bigg[1+{ \frac{ 2{H}x_g\delta \cos(\Theta + \sigma)}{{c_f}\cos\sigma(H-r_s \sin \Theta)}}\Bigg]
\end{equation}
When we plug Eq.~(25) into Eq.~(3), we find
\begin{equation}
\begin{aligned}
	\Bigg[x_g\frac{\cos(\Theta + \sigma)}{\cos\sigma}\Bigg]^2 + y_g^2 
	= y_R^2 + 
	2\frac{ x_g\cos(\Theta + \sigma) }{\cos\sigma} 
	\\\cdot
	\frac{{H}\delta y_R^2}{{c_f}(H-r_s \sin \Theta)}
	\end{aligned}
\end{equation}
After some rearrangement of terms, Eq.~(26) turns into the following:
\begin{equation}
\begin{aligned}
	\Bigg[x_g - \frac{ y_R^2 ~\delta ~H\cos\sigma}
	{{c_f}(H-r_s \sin \Theta)\cos(\Theta+\sigma)}\Bigg]^2
	\frac{\cos^2(\Theta + \sigma)}{\cos^2\sigma} + y_g^2 
	\\= y_R^2 \Bigg[1+\Bigg( \frac{y_R ~\delta ~H }{{c_{f}(H-r_s \sin \Theta})} \Bigg) ^2 \Bigg]
\end{aligned}
\end{equation}
The modeling predicts that for the situation $\Phi=0^o$, the centre of equidensity ellipses will shift along the positive x-axis in the observation plane on the ground, as seen below when the above equation is compared to Eq.~(5).
\begin{equation}
	x_C \cong \frac{ y_R^2 ~\delta ~H\cos\sigma}{{c_f}(H-r_s \sin \Theta)\cos(\Theta+\sigma)}
\end{equation}
Now inserting the parametric form of $\delta$ here, we then have,
\begin{equation}
	x_C = H{c_f}^{-1} y_R^{2-\kappa} r_0^\kappa \eta (\alpha \kappa)^{-1}
	\frac{\tan \Theta}{\cos(\Theta+\sigma)} 
	\cdot
	\frac{\cos\sigma}{H-r_s \sin \Theta}
\end{equation}
Here, it is evident that the attenuation of EAS particles shifts the centre of the equidensity ellipse towards the early part of the EAS since $x_{C} > 0$.  

\subsection{The polar density distribution of EAS particles: An elliptic lateral density function}
Eq.~(29) provides $x_C$ anticipated by the model. Additionally, it can be observed that $x_C$ and $y_R$ are related by an exponent $(2-\kappa)$. The following section uses MC data from two high-energy hadronic interaction models to display $x_C$ against $y_R$. A function almost identical to the one predicted by the parametrization is used to fit the data points. Let us rewrite Eq.~(29) in a more compact form by applying it to fit the MC data. 
\begin{equation}
	x_C = A ~y_R^B~ tan \Theta
\end{equation}
We get $B$ as close to 1 for the fit parameters in Table I below using simulated LDD data of $e$-s. It is important to note that muon data produce results comparable to those displayed here.\\
\begin{table}[!h]
 \renewcommand*{\arraystretch}{1.5}
  \begin{center}
   \begin{tabular}{|c|c|c|c|c|}
	\hline
	Species	& $~E$ (PeV) & $~~\Theta~~$  & $~A~$ & $~B~$ \\
	\hline
	Fe	& 100 	& $50^o$ & 0.017 & 1.29 \\
	\hline
	p	& 100 	& $50^o$ & 0.012  & 1.27 \\
	\hline
\end{tabular}
\caption{Analysis of simulated LDD data of $e$-s yielded $A$ and $B$ parameters through $\chi^{2}$ fits.}
 \end{center}
  \end{table}

The equation for $x_C$ can be rewritten as follows to obtain the new form that corresponds to $B\approx 1$: 
\begin{equation}
	x_C = 2 A_c y_R \tan{\Theta},
\end{equation}
where $A_c$ stands for

\begin{equation}
	A_c = H{c_f}^{-1} r_0^\kappa \eta (\alpha \kappa)^{-1}
	\cdot
	\frac{\cos\sigma}{2\cos(\Theta+\sigma)(H-r_s \sin \Theta)}
\end{equation}
Thus, Eq.~(7) becomes, 
\begin{equation}
	(x_g^2-4 x_g \cdot A_c y_R\cdot \tan\Theta)\cdot \frac{\cos^2(\Theta + \sigma)}{\cos^2\sigma} + y_g^2 = y_R^2
\end{equation}

\begin{equation}
\begin{aligned}
	y_R^2 + 4 x_g \cdot A_c y_R\cdot \tan\Theta \cdot \frac{\cos^2(\Theta + \sigma)}{\cos^2\sigma}\\ - \Bigg(y_g^2 + x_g^2 \frac{\cos^2(\Theta + \sigma)}{\cos^2\sigma}\Bigg)= 0
\end{aligned}
\end{equation}
After solving the abovementioned equation, we obtain the expression for $y_R$.
\begin{equation}
\begin{aligned}
	y_R = & -2 A_c x_g \tan\Theta \cdot \frac{\cos^2(\Theta + \sigma)}{\cos^2\sigma}
       + \Bigg[ y_g^2 + x_g^2
	   \\ & \cdot \frac{\cos^2(\Theta + \sigma)}{\cos^2\sigma} \Bigg (1+4A_c^2\tan^2\Theta  \cdot \frac{\cos^2(\Theta + \sigma)}{\cos^2\sigma}\Bigg) \Bigg]^{1/2}
\end{aligned}
\end{equation}	

\begin{equation}
y_R \approx -2 A_c x_g \tan\Theta \frac{\cos^2(\Theta + \sigma)}{\cos^2\sigma}
+\Bigg(y_g^2 + x_g^2 \frac{\cos^2(\Theta + \sigma)}{\cos^2\sigma} \Bigg)^{1/2}
\end{equation}	
If $A _c<<1$ and it has been substantiated by the simulated data that $A _c$ $\sim 10^{-2}$, we then obtain the Eq.~(36) from Eq.~(35). Eq.~(36) may be transformed into the following form by using polar coordinates, $(r_g,\beta_g)$, in place of the set $(x_g,y_g)$ i.e. $x_g = r_g \cos\beta_g$ and $y_g = r_g \sin\beta_g$ 
\begin{equation*}
\begin{aligned}
y_R \approx &
-2 A_c r_g \cos\beta_g \tan\Theta \cdot \frac{\cos^2(\Theta +\sigma)}{\cos^2\sigma}
\\& +r_g \Bigg(\sin^2\beta_g + \cos^2\beta_g \frac{\cos^2(\Theta + \sigma)}{\cos^2\sigma} \Bigg)^{1/2}
\\ \approx &
-2 A_c r_g \cos\beta_g \tan\Theta \cdot \frac{\cos^2(\Theta + \sigma)}{\cos^2\sigma}
\\ & +r_g \Bigg[1-\cos^2\beta_g\Bigg(1 - \frac{\cos^2(\Theta + \sigma)}{\cos^2\sigma}\Bigg) \Bigg]^{1/2}
\end{aligned}
\end{equation*}
Due to the very narrow opening angle of the shower cone, $\cos^2\sigma \approx 1$. Consequently, $y_R$ gets its final expression as
\begin{equation}
\begin{aligned}
	y_R \approx & -2 A_c\cdot r_g \cos\beta_g \tan\Theta \cos^2(\Theta + \sigma)
	\\& + r_g\Bigg(1-\cos^2\beta_g \sin^2(\Theta+\sigma)\Bigg)^\frac{1}{2}
\end{aligned}
\end{equation}
The second component of Eq.~(37) emerges solely from the geometric effect, whereas the first term is the outcome of the attenuation process. 
           
Based on some approximations in cascade theory [1-3], the solution of the 3D diffusion equations can yield the SLDF of cascade particles via the well-known NKG structure function [2], which is given by 
\begin{equation}
\rho(r_s)=C(s_\perp)N_e \cdot (r_s/r_0)^{s_\perp-2} (1+r_s/r_0)^{s_\perp-4.5},
\end{equation}
where, $C(s_\perp)=\frac{\Gamma(4.5-s_\perp)}{2\pi r_0^2\Gamma(s_\perp)\Gamma(4.5-2s_\perp)}$ 
usually acts as a normalization factor, and $r_0$ is called the Moliere radius.

By replacing the variable $r_s$ in Eq.~(38) with $y_R$, the ELDF for the polar density distribution (PDD) of EAS particles in the observation plane on the ground can be derived from the NKG type SLDF. According to the \textit{cone model}, the density of $e/\mu$-s can therefore be characterized by an ELDF with the following form: 
\begin{equation}
\rho(r_g,\beta_g)=\cos\Theta\cdot C(s_\perp)N_e \cdot (y_R/r_0)^{s_\perp-2} (1+y_R/r_0)^{s_\perp-4.5},
\end{equation}
where $y_R$ follows Eq.~(37).

\section{Monte Carlo simulation of cosmic-ray showers}

EAS events are simulated in the framework of the air shower simulation code  CORSIKA ver. 7.690 to obtain the $e/\mu$ LDDs/PDDs [23]. Two distinct models, QGSJet 01 ver. 1c [24] and EPOS-LHC [25] have been adopted in treating the high-energy (above 80~GeV/n) hadronic interactions. Each high-energy model is embedded with the low-energy (below 80~GeV/n) hadronic interaction model UrQMD [26]. The EGS4 [27] program library, which involves all of the significant interactions of electrons and photons, is used to simulate the EM component of an EAS.
  
At the location of the KASCADE site, the EAS events have been simulated (latitude $49.1^o$ N, longitude $8.4^o$ E, $110$ m a.s.l.) [28]. For CR secondaries such as $\mu$-s, and $e$-s, the kinetic energy cut-offs are set at $0.3$, and $0.003$ GeV, respectively. The MC showers are generated for proton (p) and iron (Fe) primaries at primary energies 5, 10, 50, 22.5, 100, 225, 500, and 1000 PeV, respectively, at $\Theta=50^o$. Additionally, we have generated some number of p- and Fe-induced showers corresponding to $\Theta$: $40^o$, $45^o$, $50^o$, $55^o$, and $60^o$ with a fixed 100 PeV energy. Furthermore, a few EAS events are generated by turning off Earth's magnetic field to observe how the geomagnetic field affects the PDDS of the hard muonic component of EASs. A summary of the MC data set and the simulation settings is given in Table II. 

\begin{table*}
\begin{center}
	\begin{tabular}{|p{1cm}p{1cm}|p{1cm}p{1cm}|p{1cm}p{1cm}|}
		\hline
		\multicolumn{1}{|p{1.3cm}|}{\multirow{2}{*}{E (PeV)}} & \multirow{2}{*}{$\Theta$} & \multicolumn{2}{p{3.5cm}|}{Event Nos: QGSJet}    & \multicolumn{2}{p{3.5cm}|}{Event Nos: EPOS-LHC}    \\ \cline{3-6} 
		\multicolumn{1}{|l|}{}                  &      & \multicolumn{1}{p{1.5cm}|}{p} & Fe & \multicolumn{1}{p{1.5cm}|}{p} & Fe \\ \hline
		\multicolumn{1}{|l|}{5}                 &  $50^o$  & \multicolumn{1}{l|}{30} & 30 & \multicolumn{1}{l|}{30} & 30 \\ \hline
		\multicolumn{1}{|l|}{10}                &  $50^o$  & \multicolumn{1}{l|}{30} & 30 & \multicolumn{1}{l|}{30} & 30 \\ \hline
		\multicolumn{1}{|l|}{22.5}              &  $50^o$  & \multicolumn{1}{l|}{25} & 25 & \multicolumn{1}{l|}{25} & 25 \\ \hline
		\multicolumn{1}{|l|}{50}                &  $50^o$  & \multicolumn{1}{l|}{25} & 25 & \multicolumn{1}{l|}{25} & 25 \\ \hline
		\multicolumn{1}{|l|}{\multirow{5}{*}{100}} &$40^o$ & \multicolumn{1}{l|}{20} & 20 & \multicolumn{1}{l|}{20} & 20 \\ \cline{2-6} 
		\multicolumn{1}{|l|}{}                  &  $45^o$  & \multicolumn{1}{l|}{20} & 20 & \multicolumn{1}{l|}{20} & 20 \\ \cline{2-6} 
		\multicolumn{1}{|l|}{}                  &  $50^o$  & \multicolumn{1}{l|}{20} & 20 & \multicolumn{1}{l|}{20} & 20 \\ \cline{2-6} 
		\multicolumn{1}{|l|}{}                  &  $55^o$  & \multicolumn{1}{l|}{20} & 20 & \multicolumn{1}{l|}{20} & 20 \\ \cline{2-6} 
		\multicolumn{1}{|l|}{}                  &  $60^o$  & \multicolumn{1}{l|}{20} & 20 & \multicolumn{1}{l|}{20} & 20 \\ \hline
		\multicolumn{1}{|l|}{225}               &  $50^o$  & \multicolumn{1}{l|}{20} & 20 & \multicolumn{1}{l|}{20} & 20 \\ \hline
		\multicolumn{1}{|l|}{500}               &  $50^o$  & \multicolumn{1}{l|}{15} & 15 & \multicolumn{1}{l|}{15} & 15 \\ \hline	
		\multicolumn{1}{|l|}{1000}              &  $50^o$  & \multicolumn{1}{l|}{10} & 10 & \multicolumn{1}{l|}{10} & 10  \\ \hline
		\multicolumn{2}{|l|}{ Total Event Nos}     & \multicolumn{1}{l|}{255} & 255 & \multicolumn{1}{l|}{255} & 255  \\ \hline
	\end{tabular}
	\caption{The Monte Carlo data set contains $1020$ of simulated p and Fe showers generated by QGSJet01~ver.~1c and EPOS-LHC hadronic interaction models.}
\end{center}
\end{table*}

In the simulation, the thinning option of CORSIKA has been employed for $E\geq 50$~PeV, using $10^{-6}$ as the thinning factor under the optimum weight limitation [29]. All the generated MC showers follow $\Phi = 0^o$ so that the EAS cores will only retain on the \emph{x}-axis.
\begin{figure}[!htbp]
	\centering
	\begin{minipage}[b]{0.5\textwidth}
	  \includegraphics[trim=0.6cm 0.6cm 0.0cm 0.6cm, scale=0.8]{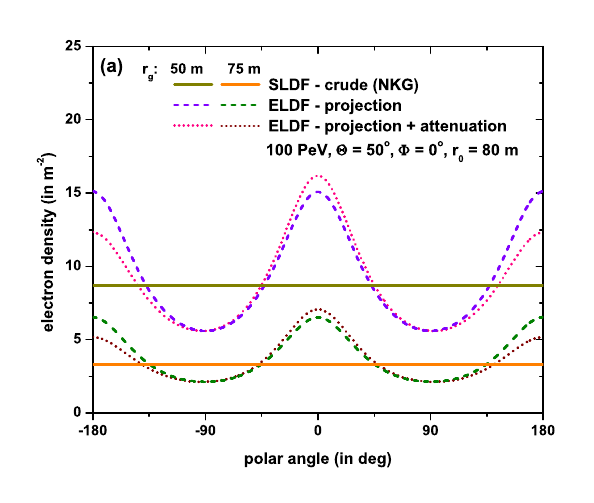}
	\end{minipage}
	\hfill
	\begin{minipage}[b]{0.5\textwidth}
	  \includegraphics[trim=0.6cm 0.6cm 0.0cm 0.6cm, scale=0.8]{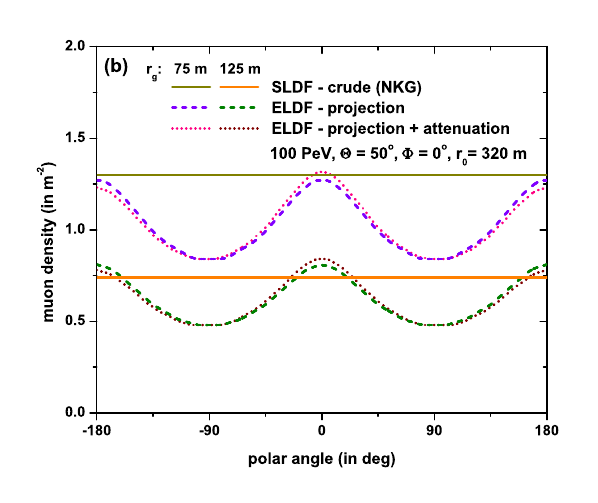}
	\end{minipage}
	\caption{Binned predicted polar variation of $e/\mu$ density of p-initiated showers at a few arbitrary core distances.}	
\end{figure}

\section{Analysis of simulated shower data}

First, we will investigate the expected behaviour of different analytical LDFs (e.g., SLDF, ELDF$-$projection, and ELDF$-$projection$+$attenuation) by studying $\rho_{g;e/\mu}$ versus $\beta_g$. These LDFs correspond Eq.~(38) with $r_{s}=r_g$ (SLDF), Eq.~(39) with the 2nd part of Eq.~(37) only for $y_R$ (ELDF$-$projection) and Eq.~(39) with the entire Eq.~(37) for $y_R$ (ELDF$-$projection$+$attenuation) respectively. Values for $s_{\perp}$, $N_e$, $N_{\mu}$, $\Theta$ in those LDFs were taken from an average 100 PeV simulated p shower with $\Theta = 50^o$ and $\Phi = 0^o $. We have calculated the analytical polar densities of $e/\mu$-s at a few arbitrarily selected core distances. For $e$-s, these distances are $r_{g}=50$~m and $75$~m, while for $\mu$-s, they are $75$~m and $125$~m. Fig. 4 shows all these variations of polar density along with polar angle. The figures reveal that compared to ELDF$-$projection and SLDF, the ELDF$-$projection $+$ attenuation offers a larger angular variation of $e/\mu$ densities. These results show how these polar variations become more pronounced at farther core distances. In Fig. 4a, the ratio between the highest and lowest analytical $e$ densities are $\frac{\rho_{r_g=50}(\beta_g=0^o)}{\rho_{r_g=50}(\beta_g=90^o)}=2.90$, and $\frac{\rho_{r_g=75}(\beta_g=0^o)}{\rho_{r_g=75}(\beta_g=90^o)}=3.35$ respectively in case of ELDF$-$projection$+$attenuation. The features remain the same for $\mu$-s as well ($\frac{\rho_{r_g=75}(\beta_g=0^o)}{\rho_{r_g=75}(\beta_g=90^o)}=1.57$ and $\frac{\rho_{r_g=125}(\beta_g=0^o)}{\rho_{r_g=125}(\beta_g=90^o)}=1.77$), which are depicted in Fig. 4b.
\begin{figure*}[!htbp]
    	\subfigure
	{\includegraphics[trim=0.6cm 0.6cm 0.6cm 0.6cm, scale=0.8]{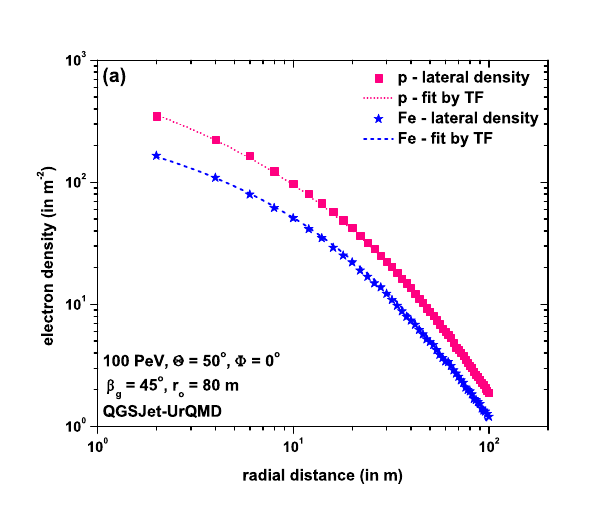}}
        \subfigure
	{\includegraphics[trim=0.6cm 0.6cm 0.6cm 0.6cm, scale=0.8]{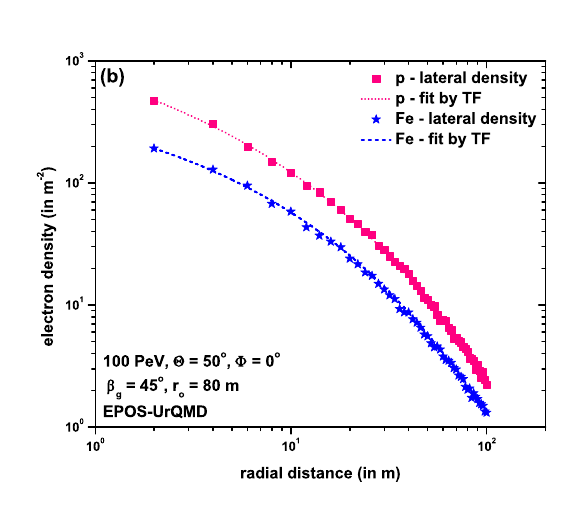}}
\\       \subfigure 
	{\includegraphics[trim=0.6cm 0.6cm 0.6cm 0.6cm, scale=0.8]{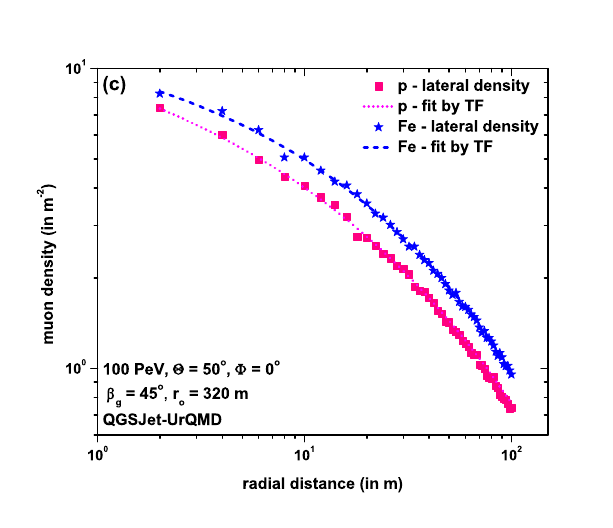}} 
         \subfigure
	{\includegraphics[trim=0.6cm 0.6cm 0.6cm 0.6cm, scale=0.8]{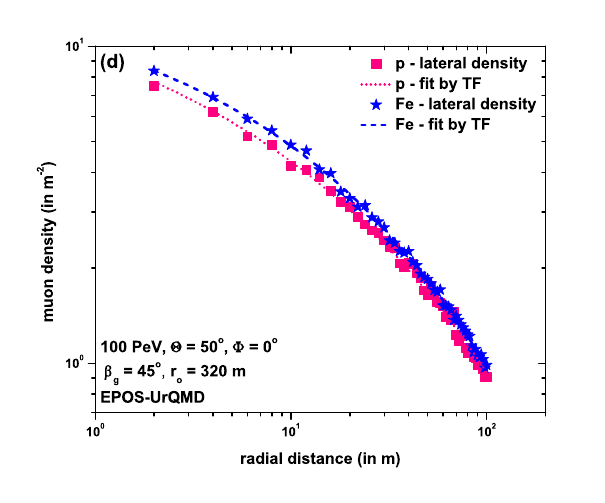}}
	\caption{Electron and muon lateral densities from simulated p- and Fe-initiated showers. Density data are fitted by the toy function (TF).}
\end{figure*}
It is evident that $x_C$, i.e. Eq.~(20) can be obtained analytically by applying a TF as a hypothesized LDF for the LDDs of $e/\mu$-s. 

We have estimated the lateral $e$ densities centred at $\beta_g = 45^o$ by taking the average in the polar region between $\beta_g = 43^o$ and $\beta_g = 47^o$ apart for simulated p and Fe showers, and their radial variations are depicted in Fig. 5a and b respectively. Similar investigations are also conducted employing $\mu$-s, and the outcomes are shown in Fig. 5c and d. The fit parameters ($\alpha$ , $\kappa$) for the models QGSJet and EPOS-LHC, which were obtained by fitting the pair of curves in Fig. 5a and b and in Fig. 5c and d by the TF, are listed in Table III. Wherever required, our studies presented in the work take the commonly accepted values for the Moliere radius $r_0$, such as $80$~m for the LDD of $e$-s and $320$~m for $\mu$-s, employed in several sea level EAS experiments [28,30].

\subsection{Estimation of the gap length ($x_C$) from the equidensity contours of electrons/muons}

Using positional information $(x_g , y_g)$ of each $e/\mu$ from the simulated data, one can easily obtain the corresponding polar coordinates ($r_g, \beta_g$) in the observational plane on the ground. We will search for the $x_C$ parameter in the LDD/PDD of $e/\mu$-s by analyzing the simulated $e/\mu$ distributions in the 2-dimensional space; $r_g - \beta_g$. The application of the method to $e$ content is expected to be slightly different from the $\mu$ content because $\mu$-s follow a different attenuation property than $e$-s. This attenuation feature is already looked after through the parameter $\Lambda$ in Sect. IIB. In the modeling for the $x_C$ and ELDF, we have not considered the effect of the geomagnetic field on EAS secondaries. Thus, it is appropriate that one should not compare model predictions with the simulation results exclusively for $\mu$-s with $B \neq 0$, i.e. $B=B_{\text{KAS}}$.    

We run the simulations using the KASCADE magnetic field, $B=B_{\text{KAS}}$ for the polar distributions of $e$-s. $B\approx 0$ was included in the simulation settings to produce most of the results involving $\mu$-s. However, we have generated a few Fe-initiated showers with $B\approx 0$, as well as $B=B_{\text{KAS}}$ to investigate the effect of the geomagnetic field on the equidensity contours of muons. It is expected that the estimated $x_C$ from the equidensity contours of $e$-s with $B=B_{\text{KAS}}$ is anticipated to be closer to the model prediction. We may expect the $\mu$-s to follow the above trend better with the case, $B\approx 0$ only in the simulation.

\begin{table*}
	\begin{center}
		\begin{tabular}
			{|l|l|l|l|l|l|l|} \hline			
			{\rm{Species}} & {\rm{Model}} & Density&  $r_{0}$ (m) & $\alpha$ & $\kappa$  & $c$  \\ \hline			
			p  & QGSJet   & $e$   & $80$   & $6.19 \pm 0.01$   & $0.35 \pm 0.0$   & $1945.6 \pm 20.2$    \\ 
			Fe &          &       &        & $5.81 \pm 0.01$   & $0.42 \pm 0.0$   & $571.2 \pm 8.5$     \\  	\hline			
			p  & EPOS-LHC & $e$   & $80$   & $6.53 \pm 0.05$   & $0.33 \pm 0.01$   & $3276.8 \pm 243.9$    \\ 
			Fe &          &       &        & $5.97 \pm 0.03$   & $0.39 \pm 0.01$   & $703.7 \pm 40.2$   \\ 			\hline 		
			p  & QGSJet   & $\mu$ & $320$  & $4.59 \pm 0.02$   & $0.32 \pm 0.01$   & $17.0 \pm 0.7$    \\ 
			Fe &          &       &        & $4.33 \pm 0.04$   & $0.39 \pm 0.01$   & $15.3 \pm 0.6 $     \\ 	\hline			
			p  & EPOS-LHC & $\mu$ & $320$  & $4.31 \pm 0.03$   & $0.31 \pm 0.01$   & $18.8 \pm 0.7$    \\ 
			Fe &          &       &        & $4.32 \pm 0.02$   & $0.36 \pm 0.01$   & $16.8 \pm 0.6$   \\ 
			\hline   			
		\end{tabular}
	\caption {Values for $\alpha$, $\kappa$ and $c$ were determined by the TF's fitting of the LDD of electrons and muons. We have used simulated p and Fe showers with $E = 100$~PeV and $\Theta = 50^{o}$.}
	\end{center}
\end{table*}    

To investigate the $x_C$ using the LDD data of $e$-s, we have chosen some p-initiated showers with $E=100$~PeV, $\Theta = 50^o$, and $\Phi = 0^o$ from the generated shower library. An equidensity contour (ellipse) for the density $\langle\rho_e\rangle \simeq 1.25$~m$^{-2}$ of $e$-s is considered and is shown in Fig. 6a corresponding to different Cartesian/polar sets of coordinates i.e. $x_g , y_g$ or $r_g , \beta_g$. The contour with a solid line in the figure represents the resulting equidensity contour having $\langle\rho_e\rangle \simeq 1.25$~m$^{-2}$ obtained from the above procedure. Applying a non-linear fit procedure to the positional data $x_g , y_g$ of $e$-s in the solid equidensity contour, an expected equidensity ellipse (dotted) is obtained. The centre of the expected elliptic contour undergoes a linear shift to a new position at C in Fig. 6a while the EAS core remains at O. Because of this, the current fit for p-initiated showers using the LDD data of $e$-s produces $x_C \simeq 9.75\pm{0.85}$~m.  

The method has also been applied to the LDD of $\mu$-s. Here, we have chosen Fe-initiated showers from the shower library. An equidensity contour (irregular solid line) for the density $\langle\rho_{\mu}\rangle \simeq 0.98~m^{-2}$ of $\mu$-s is considered corresponding to $B\approx 0$,  and is shown in Fig. 6b. The inner dashed line of Fig. 6b shows the equidensity curve with $B\neq 0$, i.e. $B=B_{\text{KAS}}$. Due to the effect of the $B$-field on $\mu$-s, there is a reduction in the semi-major axis of the inner (dashed) ellipse. The equidensity ellipse (dotted line) predicted by the fit procedure has described the LDD data of $\mu$-s very accurately corresponding to the specific case, $B\approx 0$. An $x_C \simeq 4.80\pm{1.09}$~m is found from the analysis. For $B=B_{\text{KAS}}$ (dashed line), $x_C$ takes a value $\simeq 3.19\pm{1.13}$~m instead.
\begin{figure}[!htbp]
	\centering
	\includegraphics[trim=0.6cm 0.6cm 0.6cm 0.6cm, scale=0.8]{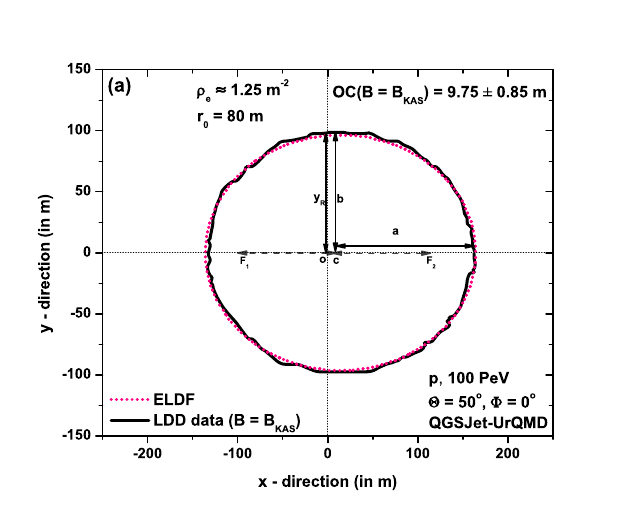}
	\includegraphics[trim=0.6cm 0.6cm 0.6cm 0.6cm, scale=0.8]{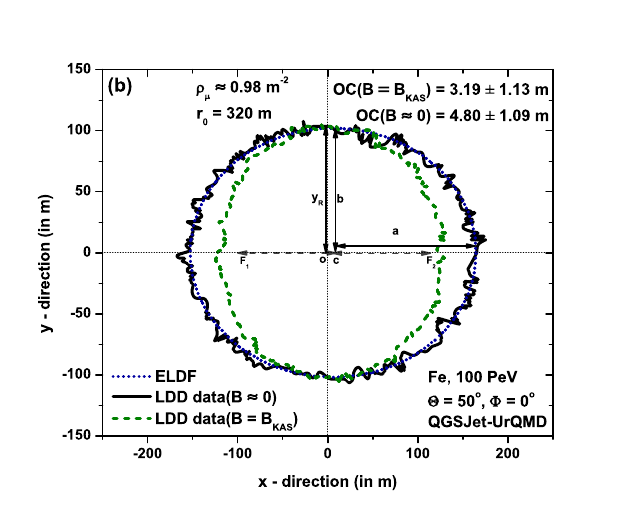}
	\caption{Formation of $x_C$ from the equidensity contours of the simulated electron (Fig. a) and muon (Fig. b) densities. $r_0$ is set at $80$~m for $e$-s and $320$~m for $\mu$-s.}
\end{figure}

\subsection{Compatibility of the gap length parameter with the analytical parametrization}

A more general analysis of lateral/polar densities of MC showers initiated by the species p and Fe at different fixed primary energies and zenith angles has been worked out. For each of the above situations, the $x_C$s are investigated with $B=B_{\text{KAS}}$ for the polar densities of $e$ from the fit procedure described in Sect. IV A. For $\mu$ densities, a similar investigation has been done for simulated showers with $B\approx 0$ instead. The model prediction for the $x_C$s according to the present parametrization using Eq.~(29) is calculated for all the showers. For $e$- and $\mu$-densities, the $x_C$s are determined for the twelve suitably chosen equidensities from the analyzed LDD data. The chosen equidensity ranges for a specific case with $\Theta = 50^o$, $E = 100$~PeV, and Fe-initiated showers are $\rho_{e}: 16.87_{(r_{g}=25~m)} - 0.11_{(r_{g}=300~m)}$~m$^{-2}$ and $\rho_{\mu}: 3.05_{(r_{g}=25~m)} - 0.22_{(r_{g}=300~m)}$~m$^{-2}$ for $e$-s and $\mu$-s respectively.

The model prediction for the expression of $x_C$ is already given through Eq.~(29). By the best fitting of LDD data of $e/\mu$-s with the TF, the values of $\alpha$ and $\kappa$ are determined. The values of $\eta$ can be found from the results in [12]. At the KASCADE location [28], the attenuation length is estimated to be $\simeq 190$~g~cm$^{-2}$ for $e$-s and $\simeq 900$~g~cm$^{-2}$ for $\mu$-s. Eq.~(29) has predicted a value of $x_C$ close to $6.67$~m for a specific case with $y_R \simeq 100.0$~m and $\Theta = 50^o$ for the LDD of $e$-s from p-initiated showers. In comparison, $x_C$ has yielded a value $\simeq 9.75\pm{0.85}$~m by our elliptical fit to the simulated equidensity contour with $\rho_e \simeq 1.25$~m$^{-2}$. While for muons using Fe-initiated showers, these values of $x_C$ are $\simeq 3.05$~m and $\simeq 4.80\pm{1.09}$~m respectively from the present parametrization, and the elliptical fitting with $\rho_{\mu}\simeq 0.98$~m$^{-2}$. The elliptical fit procedure uses $\chi^{2}$-minimization of the sum of squares the density contours by an ellipse of the form: ${\frac{(x-x_{C})}{a^2}}^2 + {\frac{(y-y_{C})}{b^2}}^2=1$, where $a$ and $b$ are clearly defined in Fig. 6a and b. 

\begin{figure}[!htbp]
	\centering
	\includegraphics[trim=0.6cm 0.6cm 0.6cm 0.6cm, scale=0.8]{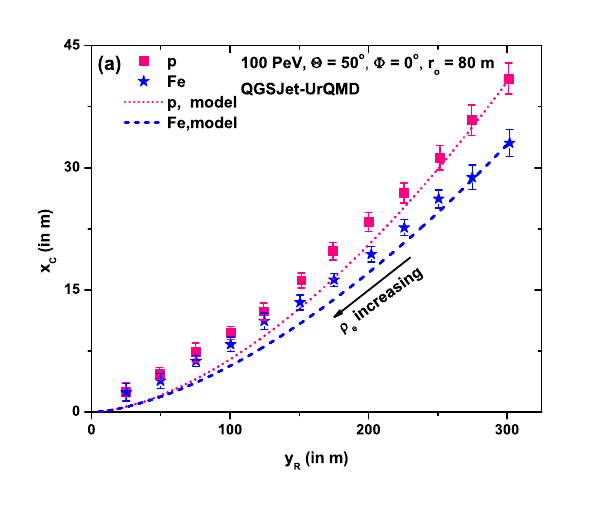}
	\includegraphics[trim=0.6cm 0.6cm 0.0cm 0.6cm, scale=0.8]{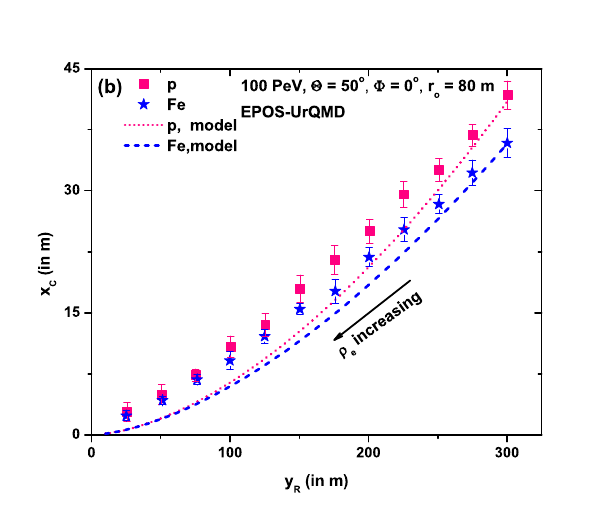}
	\caption{Correlations between $x_C$ and the variation of the $y_R$ for the electron LDD data. Dotted and dashed lines: predicted values by the Eq.~(29).}
\end{figure}

\section{Detailed simulation results on the gap length parameter}
\subsection{Basic results}

\begin{figure}[!htbp]
        \centering
	\includegraphics[trim=0.6cm 0.6cm 0.0cm 0.6cm, scale=0.8]{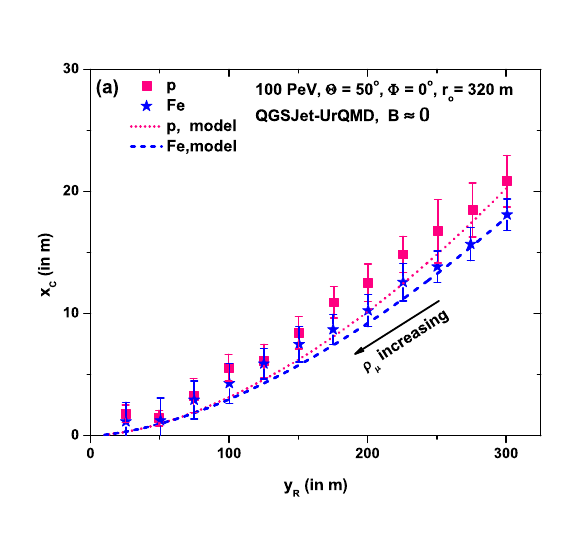}
	\includegraphics[trim=0.6cm 0.6cm 0.0cm 0.6cm, scale=0.8]{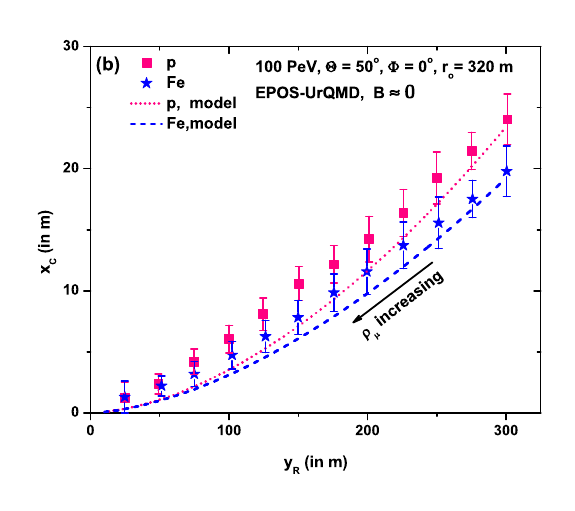}
        \caption{Correlations as in Fig. 7 but for muons. Here, the dotted and dashed lines show the predicted values of $x_C$.}
\end{figure}

Here, using $e/\mu$ LDD/PDD data obtained from MC showers initiated by p and Fe primaries and the present parametrization, the variation of $x_C$ parameter correlating with $y_R$ is thoroughly studied. Each investigation uses two different data sets consisting of twelve mean equidensities of $e$-s and $\mu$-s covering the radial distance range $25-300$~m. The main concern of the current effort is to examine the primary CR mass sensitivity of $x_C$. We will also discuss the findings on how several factors, like CR energy, zenith angle, high-energy hadronic interaction models, etc., affect $x_C$. From here onwards, our reported results based only on the fitting procedure in the paper follow $B\approx 0$ for $\mu$-s while it is just the $B=B_{KAS}$ for $e$-s.     

The correlation between $x_C$ and the variation of the $y_R$ for a fixed set of $E$, $\Theta$, $\Phi$  corresponding to QGSJet and EPOS-LHC models are demonstrated in Fig. 7a and b. Here, the LDDs of $e$-s initiated by p and Fe simulated showers are used. In Fig. 8a and b, the above correlations referring to muons for $B\approx 0$ are presented. 

\begin{figure}[!htbp]
	\centering
	\includegraphics[trim=0.6cm 0.6cm 0.0cm 0.6cm, scale=0.8]{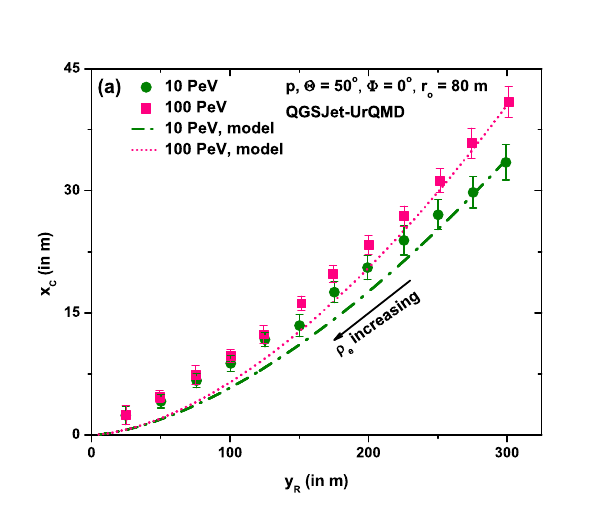}
	\includegraphics[trim=0.6cm 0.6cm 0.0cm 0.6cm, scale=0.8]{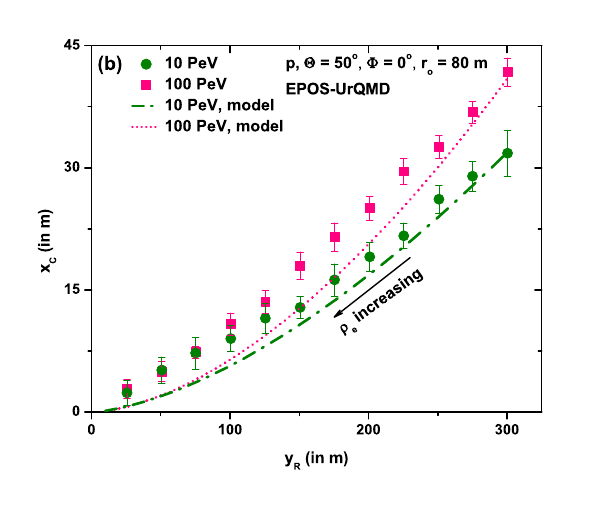}
	\caption{Correlations between $x_C$ and the variation of $y_R$ for the electron LDD data of p-initiated showers with $E=10$ and $100$~PeV.}
\end{figure}

The correlations between $x_c$ and the variation of the $y_R$ are found to be almost independent of the high-energy hadronic interaction model while dealing with the LDD of $e$-s. However, in the case of the LDDs of $\mu$-s, a slightly higher value of $x_c$ results from the EPOS-LHC model at some $y_R$  far away from the EAS core. We know that EPOS-LHC generates slightly more $\mu$-s than QGSJet [25], and these excess $\mu$-s may enhance the overall attenuation of $\mu$-s, thereby giving a slightly higher value for $x_C$. Here, it is found that the parametrization predicts $x_C$ to be very low, close to the EAS core and a steady rise with increasing $y_R$ until it hits the fitted values for $y_{R} \geq 275$~m irrespective of CR particles. The model maintains a constant height $H$ of the apex across the late and early zones of the conical shower front. However, in actuality, $H$ should depend on the EAS core distance, or $y_R$, i.e., $H(y_{R})$ [31]. It implies a high value of $H$ near the EAS core, resulting in extremely low values for $x_C$. $H$ falls with $y_R$ beyond the EAS core, providing substantially larger values for $x_C$. The discrepancy between the parametrization and the simulation may come from p-initiated showers not being attenuated in the atmosphere like Fe-initiated showers. One more probable source of discrepancy between the parametrization and the simulation may arise from the local attenuation $\eta{(y_R)}$ since the atmospheric composition may vary with $y_R$, and that might affect $\eta$ or $c_{f}$.

\begin{figure}[!htbp]
     	\centering
	\includegraphics[trim=0.6cm 0.6cm 0.0cm 0.6cm, scale=0.8]{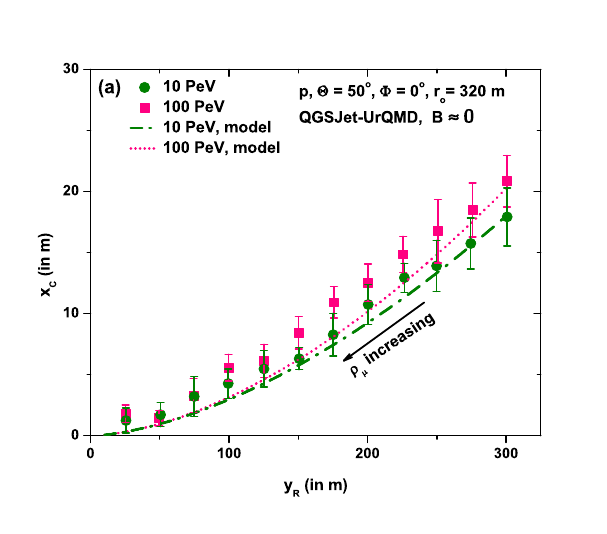}
	\includegraphics[trim=0.6cm 0.6cm 0.0cm 0.6cm, scale=0.8]{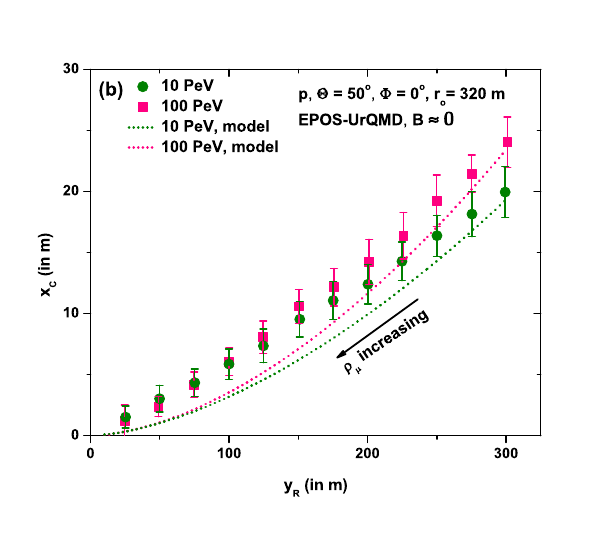}
	\caption{Correlations as in Fig. 9 but for muons. Here, the dotted and dashed lines show the predicted values of $x_C$.}
\end{figure}

Fig. 9 shows the variation of $x_C$ with $y_R$ for two given primary energies using LDD/PDD data of $e$-s of p-initiated showers. We noticed no appreciable difference from the comparison of the results based on the QGSJet model (Fig. 9a) with those from the EPOS-LHC model (Fig. 9b). It may be concluded from the analysis of the energy dependency of $x_C$ as displayed in Fig. 9a and b that $x_C$ takes higher values for higher energies, and manifest themselves so obviously for large values of $y_R$. The model predictions in  Fig. 9 exhibit a behaviour similar to that shown in Fig. 7 above. 
\begin{figure}[!htbp]
	\centering
	\includegraphics[trim=0.6cm 0.6cm 0.0cm 0.6cm, scale=0.8]{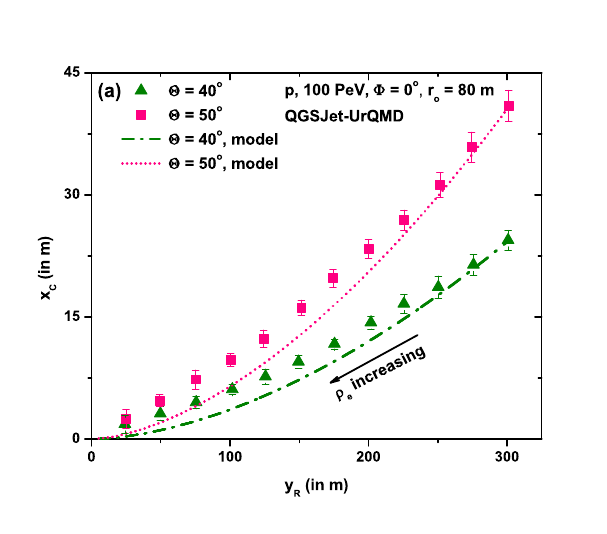}
	\includegraphics[trim=0.6cm 0.6cm 0.0cm 0.6cm, scale=0.8]{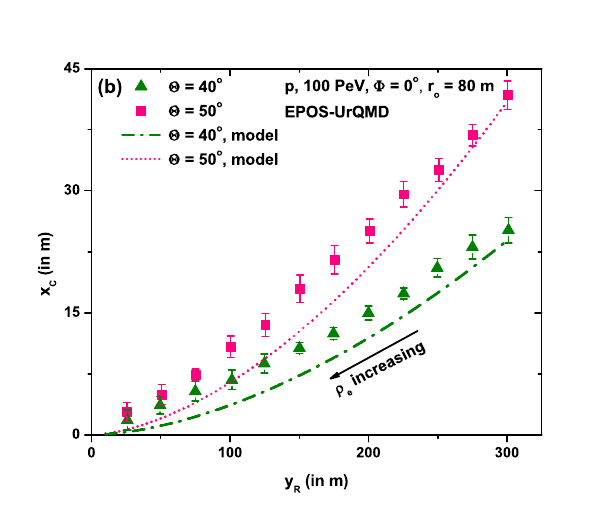}
	\caption{Correlations between $x_C$ and the variation of $y_R$ for the electron LDD data of p-initiated showers at zenith angles, $40^o$ and $50^o$.}
\end{figure} 

Similar variations of $x_C$ against $y_R$ based on the QGSJet and EPOS-LHC models using the fit method on the equidensity contours of $\mu$-s are presented through Fig. 10a and b. The results are compared with the parametrized predictions as well. The nature of variation of the $x_C$ in Fig. 10 agrees with that which is reported in Fig. 8.
\begin{figure}[!htbp]
        \centering
	\includegraphics[trim=0.6cm 0.6cm 0.0cm 0.6cm, scale=0.8]{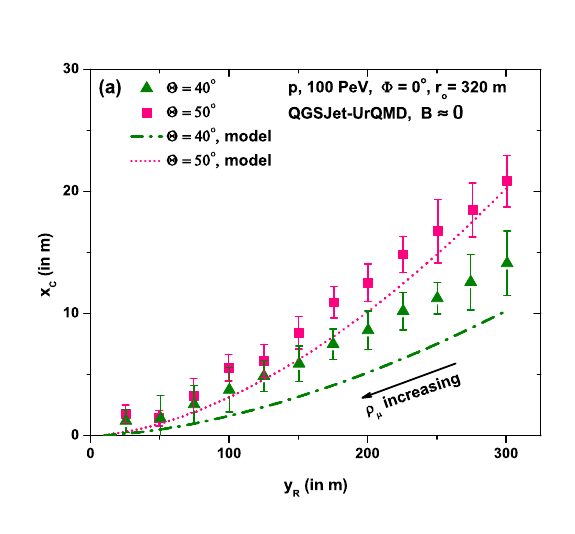}
	\includegraphics[trim=0.6cm 0.6cm 0.0cm 0.6cm, scale=0.8]{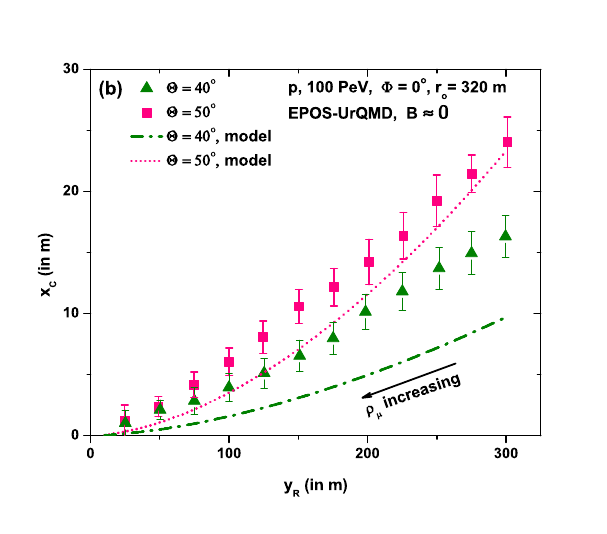}
	\caption{Correlations as in Fig. 11 but for muons. Here, the model predicted values of $x_C$ are shown by the dotted and dashed lines.}
\end{figure}

Fig. 11 presents the curves $x_C$ versus $y_R$ at two values of $\Theta$ (taking LDD/PDD data of $e$-s from p-initiated showers) corresponding to both the QGSJet and EPOS-LHC models. The results of our parametrized predictions are also included in Fig. 11a and b for $\Theta = 40^o$ and $\Theta = 50^o$, respectively. In the case of LDD/PDD data of $\mu$-s, the above studies are also carried out and are shown in Fig. 12a and b. The dotted and dashed lines in these figures show the results predicted by the parametrization. Results on $x_C$ predicted by the parametrization, particularly at $\Theta = 40^o$ in Fig. 12a and b showing a significant departure from the simulations (fit-procedure) in the regions where equidensities of $\mu$-s are very low. Although we can claim that $\Theta$ has a significant effect in Eq.~(29) for $\mu$-s, $\eta$'s effect cannot be completely ruled out ($\eta$ or $\Lambda$ for the attenuation of $e/\mu$-s in CR showers must differ). 
\subsection{Results concerning the cosmic ray mass sensitivity of $x_{C}$}
\begin{figure}[!htbp]
     	\centering
	\includegraphics[trim=0.6cm 0.6cm 0.0cm 0.6cm, scale=0.8]{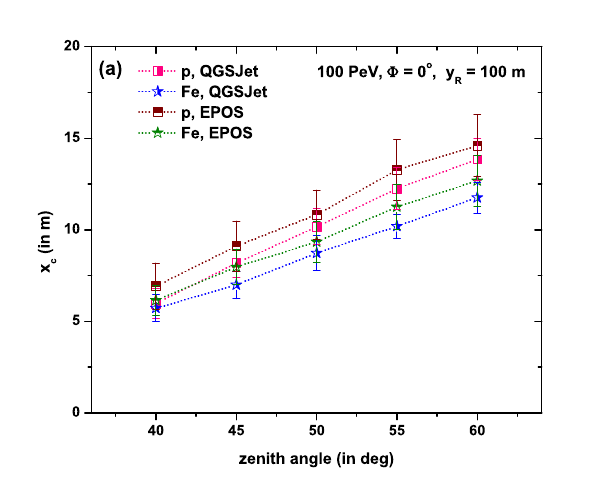}
	\includegraphics[trim=0.6cm 0.6cm 0.6cm 0.6cm, scale=0.8]{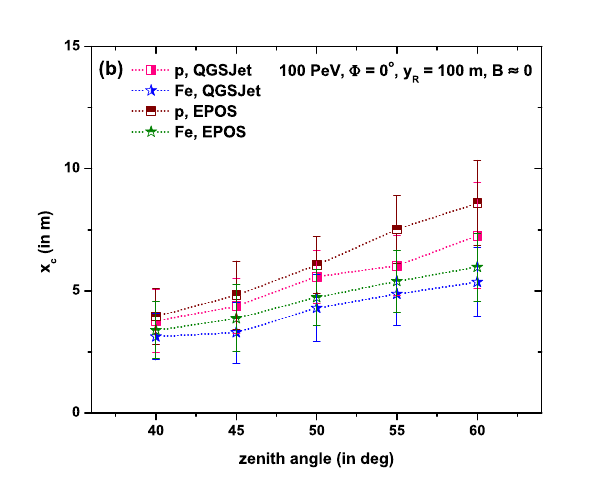}
	\caption{Average $x_C$ obtained from the equi-density fit procedure shown as a function of $\Theta$ at a fixed $y_R=100$~m for p- and Fe-initiated showers by exploiting two high-energy hadronic interaction models. Fig. a (electron data) and Fig. b (muon data).The lines are only a guide for the eye.}
\end{figure}

Now, we will look into whether or not $x_C$ exhibits sensitivity to CR mass composition. To do this, first, we must analyze its variation with $\Theta$ for the simulated p- and Fe-initiated showers either at a fixed $y_R$ or fixed $\rho_{e}$ or $\rho_{\mu}$. Another attempt is made to describe the relationship between $x_C$ and the variation of CR energy $E$ at a constant $\rho_{e}$ or $\rho_{\mu}$.\\ 

Fig. 13 and Fig. 14 display the studies on $x_C$ with respect to the variation of $\Theta$ for a fixed value of $y_R$ and $\rho_{e}$ or $\rho_{\mu}$ repectively. Results shown in Fig. 13a are based on the LDD/PDD data of $e$-s while Fig. 13b relate with the $\mu$ LDD/PDD data. Moreover, the dependency of $x_{C}$ on the hadronic interaction models, QGSJet and EPOS-LHC, if any, has been put into Fig. 13a and b. All these studies on $x_C$ versus $\Theta$ variations are repeated in Fig. 14 while keeping $\rho_{e}$ or $\rho_{\mu}$ at certain fixed values. It can be understood from Fig. 13 that for a fixed $y_R$, $x_C$ assumes greater values with $\Theta$ regardless of the CR species and the hadronic interaction models. Otherwise, this suggests a substantial correlation between $x_C$ and the atmospheric slant depth.

It is worth mentioning that Fig. 14 is an equivalent representation of  Fig. 13 where $x_C$ decreases with increasing $\Theta$ instead. Highly inclined showers are expected to suffer more attenuation in the atmosphere than nearly vertical showers. Consequently, a particular $e/\mu$ density occurs at a lower core distance for a highly inclined shower than a nearly vertical shower. Hence, $x_C$ would be smaller for highly inclined showers (say, $\Theta = 60^o$) than low zenith showers corresponding to a fixed $e/\mu$ density.    

\begin{figure}[ht]
        \centering
	\includegraphics[trim=0.6cm 0.6cm 0.0cm 0.6cm, scale=0.8]{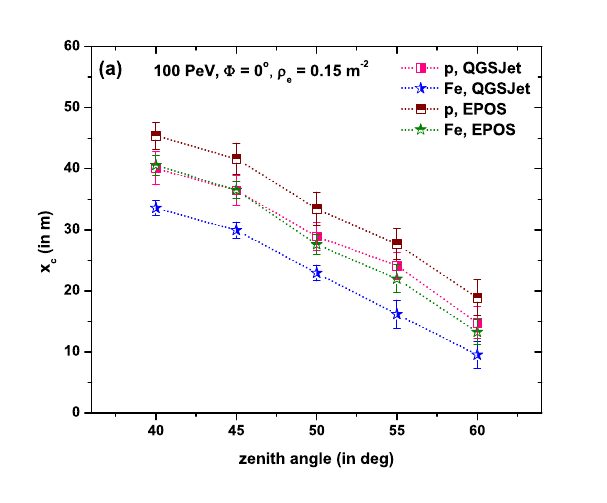}
	\includegraphics[trim=0.6cm 0.6cm 0.6cm 0.6cm, scale=0.8]{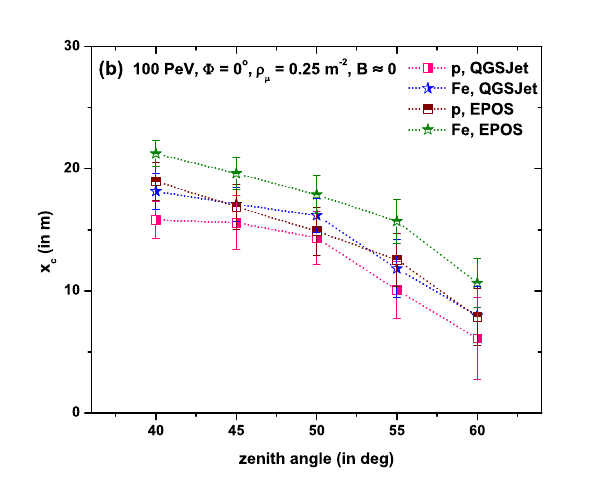}
\caption{Average $x_C$ obtained from the equi-density fit procedure shown as a function of $\Theta$ at fixed $\rho_e$ (Fig. a) and $\rho_\mu$ (Fig. b) for simulated p and Fe showers by exploiting two high-energy hadronic interaction models.The lines are only a guide for the eye.}
\end{figure}
Fig. 15a illustrates a correlation between $x_C$ and $E$ and its model dependencies if any, for a given $\Theta = 50^o$ and $\rho_{e} \simeq 1.5$~m$^{-2}$ using simulated p and Fe showers. The results of the same study but using the LDD/PDD of $\mu$-s with $\rho_{\mu} \simeq 1.0$~m$^{-2}$ are given in Fig. 15b. Fig. 15a reveals that $x_C$ gives higher  values for p-initiated showers than for Fe. It indicates that equidensity ellipses in p-initiated LDDs experience more stretching along the semi-major axis at a fixed $\rho_{e} \simeq 1.5$~m$^{-2}$. It is found that, given the same $\rho_e$, the area of the equidensity contour of a p-initiated shower is larger than the area of a Fe-initiated shower (it is a generic feature that a p shower contains more $e$-s than a Fe shower). However, the above feature is completely reversed when the LDD of $\mu$-s is used (Fig. 15b). In this case, p showers generate fewer muons than Fe showers. The variations in Fig. 15a and b also reveal that the EPOS-LHC model contributes higher values of $x_C$ compared to QGSJet.             
\begin{figure}[!htbp]
	\centering
	\includegraphics[trim=0.6cm 0.6cm 0.0cm 0.6cm, scale=0.8]{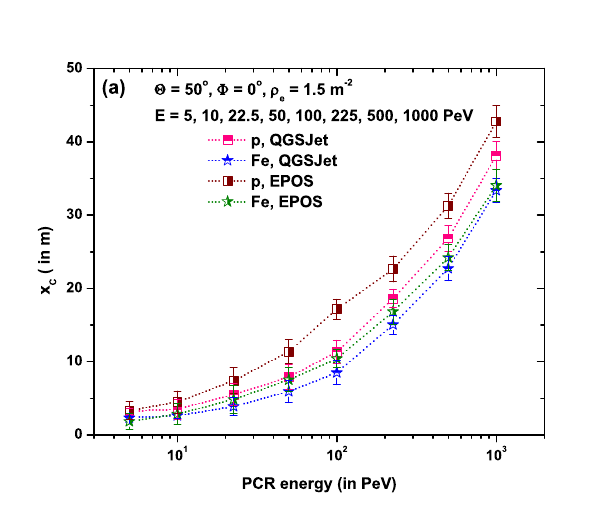}
	\includegraphics[trim=0.6cm 0.6cm 0.0cm 0.6cm, scale=0.8]{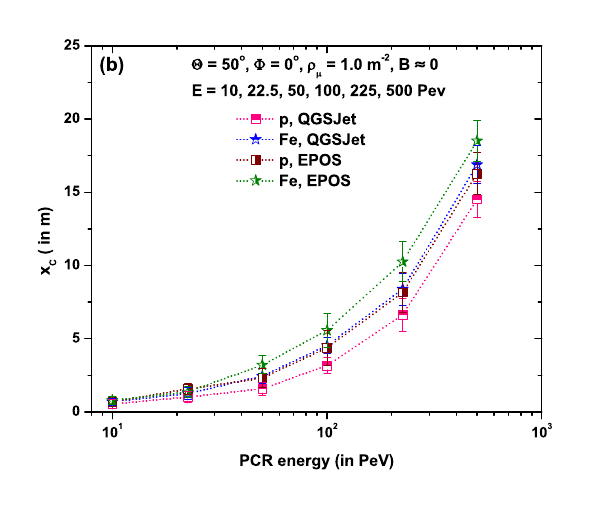}
	\caption{Average $x_C$ obtained from the equi-density fit procedure shown as a function of CR energy, $E$ at a fixed $\rho_e$ (Fig.~a), and $\rho_\mu$ (Fig.~b) for p- and Fe- initiated showers. The lines are only a guide for the eye.}
\end{figure}

\section{Comparison of simulated polar and lateral distributions of electrons and muons with the ELDF predictions}

\begin{figure}[!htbp]
	\centering
	\includegraphics[trim=0.6cm 0.6cm 0.0cm 0.6cm, scale=0.8]{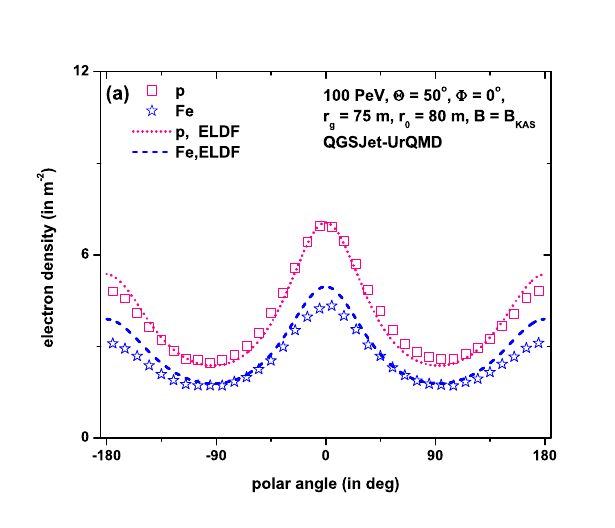}
	\includegraphics[trim=0.6cm 0.6cm 0.0cm 0.6cm, scale=0.8]{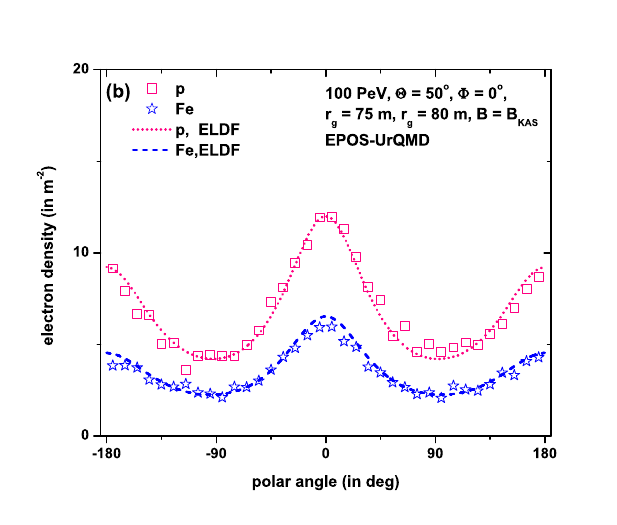}
	\caption{The variation of binned electron density as a function of polar angle $\langle{\beta_g}\rangle$ with $B = B_{KAS}$. The dotted (p) and dashed (Fe) curves stand for the fitted polar density values predicted by the ELDF.}
\end{figure}

\begin{figure}[!htbp]
	\centering
	\includegraphics[trim=0.6cm 0.6cm 0.0cm 0.6cm, scale=0.8]{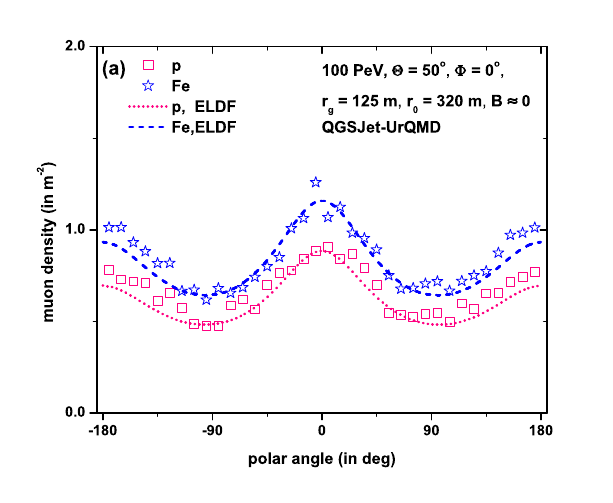}
	\includegraphics[trim=0.6cm 0.6cm 0.0cm 0.6cm, scale=0.8]{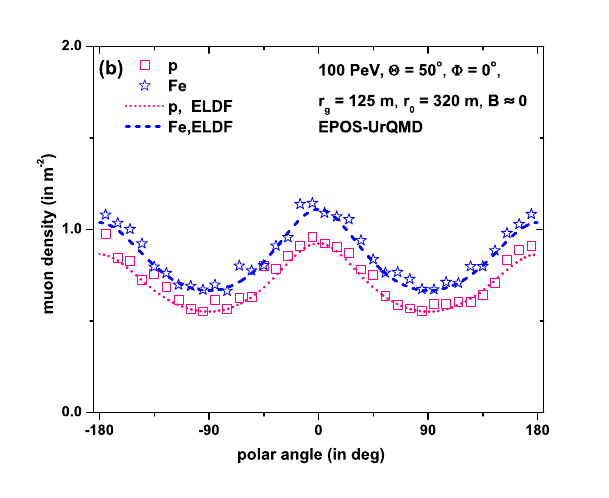}	
	\caption{Same as Fig. 16 but for muon densities with  $B\approx 0$. The dotted (for p) and dashed (for Fe) curves stand for the fitted polar density values predicted by the ELDF.}
\end{figure}

The analytical method implemented in this work has predicted the polar and lateral structures of EAS $e/\mu$-s through the ELDF in Eq.~(39), including $x_C$. The approximate form of the $y$-coordinate $y_R$, which refers to an EAS particle's observation plane coordinates ($r_g, \beta_g$) on the ground, is given in Eq.~(37). We have investigated the mean polar and lateral density variations of $e/\mu$-s of simulated p- and Fe-initiated showers together with the ELDF predictions. To obtain each of the several EAS observables, such as the expected density at a specific $r_g, \beta_g$: $\rho{(r_g, \beta_g})$ ${N_{e}}~{\text{or}}~{N_{\mu}}$, $s_{\perp}$ etc., the relevant LDFs (in this case, the ELDF) are employed through fitting of simulated densities of $e/\mu$-s.
\begin{figure}[!htbp]
	\centering
	\includegraphics[trim=0.6cm 0.6cm 0.0cm 0.6cm, scale=0.8]{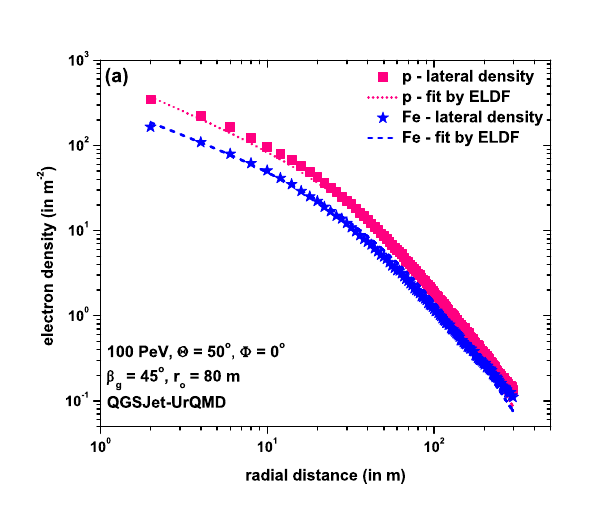}
	\includegraphics[trim=0.6cm 0.6cm 0.0cm 0.6cm, scale=0.8]{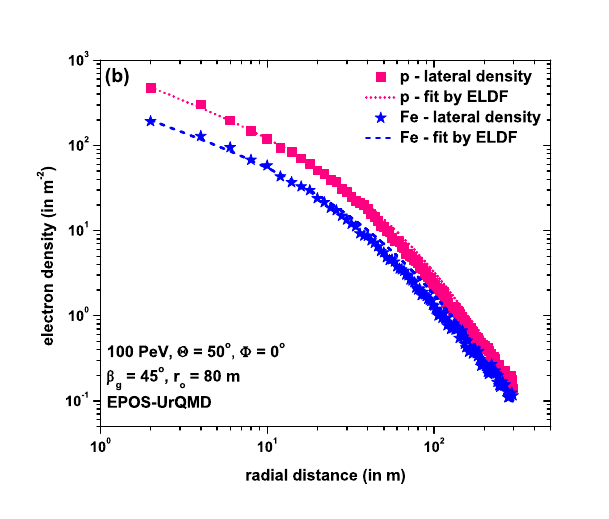}
	\includegraphics[trim=0.6cm 0.6cm 0.0cm 0.6cm, scale=0.8]{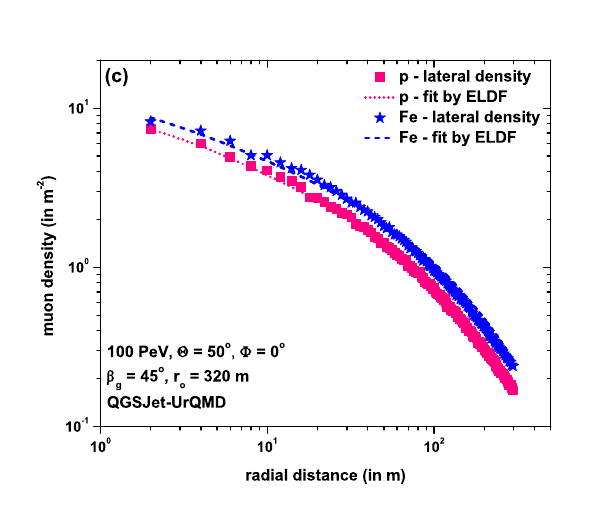}
	\includegraphics[trim=0.6cm 0.6cm 0.0cm 0.6cm, scale=0.8]{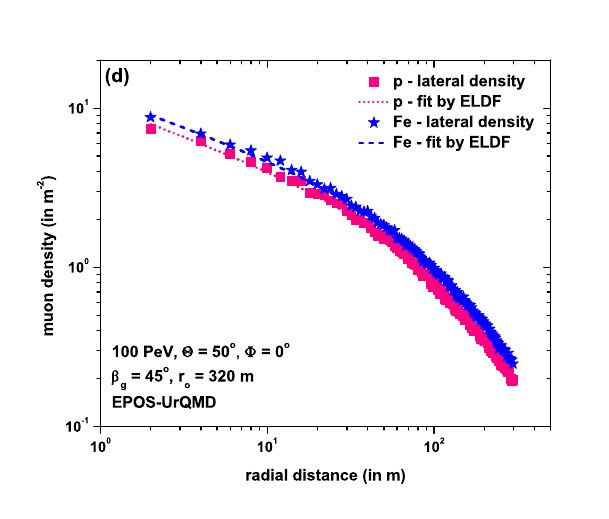}	
	\caption{The variation of binned electron/muon density as a function of radial distance. The dotted (p) and dashed (Fe) curves are the fitted density data by the ELDF.}
\end{figure}

We have examined the polar density variations of $e$-s at an arbitrarily chosen radial distance $r_g =75$~m. In the case of $\mu$-s polar densities, the radial distance is set at a slightly higher value, $125$~m, because $\mu$-s have a longer mean free path than $e$-s as they move towards the ground. Without regard to the values of $r_g$, the full angular size ($-180^{o}, 180^{o}$) is binned with bin sizes of $\Delta{\beta_g}=10^o$ each together with a radial distance bin $\Delta{r_g}=0.75$~m at about any particular $r_g$ value. Fig. 16a - b and Fig. 17a - b present our simulated results along with the dependencies of the high-energy hadronic interaction models. The same figures displayed the ELDF predictions from the fits of simulated polar densities by the ELDFs. The figures indicate that the ELDF, using the appropriate set of parameters from Table IV, could nicely approximate the average polar density variations of $e/\mu$-s of simulated p and Fe showers.

\begin{table*}
	\begin{center}
		\begin{tabular}
			{|l|l|l|l|l|l|l|l|r|} \hline
			
			{\rm{Species}}  & {\rm{Model}}  &  {\rm{Density}}  & $r_{0}$ (m) & $\alpha$ & $\kappa$  & $N_{e} {\text{or}} N_{\mu}$  & $s_{\perp}$   
			\\ 
			\hline
			\hline   
			p  & QGSJet   & $e$  & $80$   & $6.19\pm{0.01}$   & $0.35\pm{0.0}$   & $458039.8$   & $1.23\pm{0.09}$      \\ 
			Fe  &   &   &    & $5.81\pm{0.01}$   & $0.42\pm{0.0}$   & $322155.0$   & $1.32\pm{0.10}$     \\ 
			\hline      
			p  & EPOS-LHC   & $e$ & $80$   & $6.53\pm{0.05}$   & $0.33\pm{0.01}$   & $744632.9$   & $1.29\pm{0.02}$       \\ 
			Fe  &   &    &    & $5.97\pm{0.03}$   & $0.39\pm{0.01}$   & $390563.7$   & $1.35\pm{0.01}$      \\ 
			\hline
			\hline 
			p & QGSJet   & $\mu$ & $320$   & $4.59\pm{0.02}$   & $0.32\pm{0.01}$   & $241953.9$   & $1.58\pm{0.01}$    \\ 
			Fe  &   &    &    & $4.33\pm{0.04}$   & $0.39\pm{0.01}$   & $358828.2$   & $1.61\pm{0.01}$    \\ 
			\hline
			p & EPOS-LHC   & $\mu$  & $320$   & $4.31\pm{0.03}$   & $0.31\pm{0.01}$   & $429283.6$   & $1.60\pm{0.0}$     \\ 
			Fe &   &    &   & $4.32\pm{0.02}$   & $0.36\pm{0.01}$   & $531572.0$   & $1.60\pm{0.01}$  \\
			\hline 							
		\end{tabular}
		\caption {Values for $\alpha$, $\kappa$ that were determined by the TF's fitting of the LDD of electrons and muons. Fit values of EAS parameters such as $s_{\perp}$ and ${N_{e}}~{\text{or}}~{N_{\mu}}$ are also shown. We have used simulated p and Fe showers with $E = 100$~PeV and $\Theta = 50^{o}$.} 
	\end{center}
\end{table*}    

In Fig. 18, the simulated lateral densities of $e/\mu$-s estimated from our pre-specified polar angle bin $\Delta{\beta_g}=10^o$ centered on $\beta_{g}=45^o$ with radial distance bin $\Delta{r_g}=0.75$~m are displayed against $r_g$. The predicted curves from the fits of the simulated density data by the ELDF are also included in these figures. \\

The essential parameters required for the ELDF fit in Fig. 18 are given in Table IV. Our analysis found the statistical uncertainties in the ELDF fit for ${N_{e}}~{\text{or}}~{N_{\mu}}$ as $\approx \pm{9}\%$ irrespective of primary species and high-energy hadronic models. We have determined MAPE (mean absolute percentage error) to measure the accuracy of our fits by the ELDFs to the simulated lateral and polar density distributions. Our ELDF-fitted LDDs deviate from the simulated ones by up to $MAPE\simeq 7\%$ for $\mu$-s with $r_0=320$~m and $MAPE\simeq 10\%$ for $e$-s with $r_0=80$~m irrespective of models. On the contrary, the ELDF-fitted polar PDDs suffer deviations up to $MAPE\simeq 6\%$ for $\mu$-s at $r_g=125$~m (for $r_0=320$~m) and $MAPE\simeq 8\%$ for $e$-s at $r_g=75$~m (for $r_0=80$~m) respectively.
\section{Summary and conclusions} 
In this work, an extensive effort has been made for a possible analytical shape of the asymmetric polar and lateral density distributions of $e/\mu$-s in the observation plane on the ground for non-vertical showers. First, we attempt to model the shift of the centre of elliptic equidensity contours. By examining meticulously the two crucial effects associated with the positional coordinates and the attenuation of EAS particles between the shower and observation planes while a shower evolves towards the ground, the so-called $x_C$ parameter is worked out. Our analytical method uses a TF as a basic LDF to describe the LDDs of $e/\mu$-s of EAS to predict the $x_C$ parameter. Then, the ELDF dependent on $x_C$ has been obtained by considering a linear relationship between $x_C$ and $y_R$ for a given pair of $\Theta$ and $\Phi$. In the paper, however, our method has exploited the relationship with the $\Phi=0^o$ case only because the shower simulations were only conducted at $\Phi=0^o$ here. A straightforward step is necessary for the $\Phi \neq{0^o}$ case if one uses $\beta_{g}-\Phi$ instead of $\beta_{g}$ in the derivation. Finally, the desired ELDF has been obtained, first by multiplying the NKG function, represented through Eq.~(38), with $\cos\Theta$, and then adopting the final expression for $y_R$ (Eq.~(37)) in place of $r_s$ in it.

Each parametrized result obtained here was validated using the simulated p- and Fe-initiated showers at the KASCADE location. We have borrowed some EAS parameters from an average p shower simulated at $E=100$~PeV, $\Theta=50^{o}$, and $\Phi=0^{o}$ in order to identify some distinguishing characteristics among the PDDS of $e/\mu$-s predicted by the LDFs namely SLDF, ELDF: projection and ELDF:projection+attenuation. These parameters include ${N_{e}}~{\text{or}}~{N_{\mu}}$, $s_{\perp}$, $\Theta$, $r_{0}$. These distributions were shown in Fig. 4. We were able to reconstruct the simulated showers by fitting the simulated polar densities of $e/\mu$-s using the full ELDF (Eq.~(39) with $y_R$ from Eq.~(37)), which are displayed in Fig. 16 and Fig. 17. It suggests that the ELDF could accurately describe the asymmetric PDDs of $e/\mu$-s of simulated showers, including $x_C$ in $y_R$. The main goal of any EAS investigation is to fit more accurately the LDD data of $e/\mu$-s using an appropriate LDF, which has been accomplished in this paper. In this regard, our full ELDF satisfies all of these requirements through the Fig. 18. $x_C$ expressed usually in $y_R$ along with some EAS parameters as well as a set of TF fit parameters like $r_0$, $\Theta$, $\sigma$, $\eta$, $\alpha$ and $\kappa$, were used to predict $x_C$ by the present parametrization. On the other hand, the expected equidensity contour and the related $x_C$ were approximated by applying a non-linear fit approach to the positional data $x_g , y_g$ of $e/\mu$-s in the simulated equidensity contour on the $x_g - y_g$ plane (Fig. 6). Hence, these parametrized and fitted values of $x_C$, as well as their variations with some key EAS observables, were shown in various figures throughout the paper. One might be able to recognize $x_C$ as a potential CR mass-sensitive observable based on the variations in $x_C$ with $\Theta$ and $E$.

Due to the influence of the geomagnetic field ($B$), asymmetry is particularly evident in the LDDS/PDDs of $\mu$-s [5,10]. It suggests that the $B$-field should act on $\mu$-s in a shower and the new position: $B(x_g , y_g , z_{g}) \rightarrow \acute{B}(\acute{x}_g , \acute{y}_g , \acute{z}_{g})$~(Fig. 1) of a $\mu$ under consideration in the observation plane on the ground is finally inevitable. However, the necessary modeling and calculations resulting from the abovementioned effect have not been incorporated into the present study. For this reason, only the $B\approx 0$ criterion in the simulation has been used for $\mu$-s to compare the results predicted by our parametrization with the simulated ones.

We have used the LDD data on $e/\mu$-s of p and Fe showers generated by the hadronic interaction models QGSJet-01c and EPOS-LHC to estimate $x_C$ and to compare it with the parametrized predictions. Our ultimate goal is to examine whether the simulated $e/\mu$ LDDS/PDDs of p- and Fe-initiated showers generated by both models can be accurately reconstructed using the full-shape function ELDF. The shapes of the simulated PDD/LDD data (shown in Fig. 16, Fig. 17 and Fig. 18) were found to be better described by EPOS-LHC model than QGSJet-01c.

The model predicted $x_C$, and the simulated ones match better on the higher side of $y_R$ or relatively lower densities of $e/\mu$-s. These discrepancies between them become smaller for somewhat highly inclined showers. Despite certain shortcomings of the modeled $x_C$ by Eq.~(31), it gives us a solution for the desired ELDF.
 
The ELDF based on the \emph{cone model} obtained here may be more effective for reconstructing the simulated/observed LDDs/PDDs of $e/\mu$-s directly in the observation plane on the ground. To infer the mass composition and energy measurements of CRs, a  study of this kind might provide a more precise estimation for the universal EAS observables $N_e$, $N_{\mu}$, $s_{\perp}$, $s_{\rm{local}}$, $\Theta$, and $\Phi$. It is also possible to look at a parametrization that adds the geomagnetic field properties in the near future, particularly concerning EAS muons.

\section*{Data Availability Statement}
As all simulated data have been presented in the main text through figures and tables; therefore, this manuscript has no associated data information.

\section*{Acknowledgment}
Authors acknowledge the financial support from the SERB, Department of Science and Technology (Govt. of India) under the Grant no. EMR/2015/001390.  


\end{document}